\documentclass[iop, twocolumn, tighten]{emulateapj}

\graphicspath{{./}{figures/}}
\usepackage{amsmath}	
\usepackage{amssymb}	
\usepackage{float}
\newcommand{\kms}{\ensuremath{\,\mathrm{km\ s}^{-1}}}
\newcommand{\kpc}{\ensuremath{\,\mathrm{kpc}}}

\usepackage{color}
\definecolor{red}{rgb}{0.8,0.0,0.0}
\definecolor{blue}{rgb}{0.0,0.0,0.8}
\definecolor{green}{rgb}{0.0,0.5,0.0}

\begin{document}
\title{Dippers from the {\it TESS} Full-Frame Images I: The Results of the first 1 year data and Discovery of A Runaway dipper}

\email{constantinoplite@gmail.com}

\author{Tomoyuki Tajiri\altaffilmark{1}, Hajime Kawahara\altaffilmark{1,2}, Masataka Aizawa\altaffilmark{3}, Michiko S. Fujii\altaffilmark{4}, Kohei Hattori\altaffilmark{5}, Yui Kasagi\altaffilmark{6}, Takayuki Kotani\altaffilmark{7,8,6}, Kento Masuda\altaffilmark{9,10}, Munetake Momose\altaffilmark{11}, Takayuki Muto\altaffilmark{12}, Ryou Ohsawa\altaffilmark{13}, Satoshi Takita\altaffilmark{7}}
 \altaffiltext{1}{Department of Earth and Planetary Science, The University of Tokyo, 7-3-1, Hongo, Tokyo, Japan}
 \altaffiltext{2}{Research Center for the Early Universe, 
School of Science, The University of Tokyo, Tokyo 113-0033, Japan}
 \altaffiltext{3}{Department of Physics, The University of Tokyo, 7-3-1, Hongo, Tokyo, Japan}
 \altaffiltext{4}{Department of Astronomy, The University of Tokyo, 7-3-1, Hongo, Tokyo, Japan}
 \altaffiltext{5}{Department of Physics, Carnegie Mellon University, 5000 Forbes Ave, Pittsburgh, PA 15213, USA}
 \altaffiltext{6}{Department of Astronomy, School of Science, The Graduate University for Advanced Studies (SOKENDAI), 2-21-1 Osawa, Mitaka, Tokyo, Japan}
  \altaffiltext{7}{National Astronomical Observatory of Japan, 2-21-1 Osawa, Mitaka, Tokyo 181-8588, Japan}
 \altaffiltext{8}{Astrobiology center, 2-21-1 Osawa, Mitaka, Tokyo 181-8588, Japan}
  \altaffiltext{9}{Institute for Advanced Study, School of Natural Sciences, 1 Einstein Drive, Princeton, NJ 08540, USA}
  \altaffiltext{10}{Department of Earth and Space Science, Osaka University, Osaka 560-0043, Japan}
 \altaffiltext{11}{College of Science, Ibaraki University, Bunkyo 2-1-1, Mito 310-8512, Japan}
  \altaffiltext{12}{Division of Liberal Arts, Kogakuin University, 1-24-2 Nishi-Shinjyuku, Shinjyuku-ku, Tokyo 163-8677, Japan}
 \altaffiltext{13}{Institute of Astronomy, The University of Tokyo, 2-21-1, Osawa, Mitaka, Tokyo 181-0015, Japan}

\begin{abstract}
We present a comprehensive catalog of the dippers---young stellar objects that exhibit episodic dimming---derived from the one year's worth of data of Transiting Exoplanet Survey Satellite ({\it TESS}) full-frame images. In the survey, we found 35 dippers using the convolutional neural network, most of them newly discovered. Although these dippers are widely distributed over the first half-hemisphere that {\it TESS} surveyed, we identified the majority's membership with  the nearest association Scorpius--Centaurus, Velorum OB2, and nearby Orion molecular cloud complex. However, several dippers are likely to be located in the field. We also found three old dippers whose age exceeds ten million year, which is considered as the disk dissipation time. The color-color diagram indicates that these old dippers are likely to have an extreme debris disk. In particular, we found a runaway old dipper having a large three-dimensional velocity of $72$ \kms. The dippers in the field, which were probably escaping from their birth molecular clouds or were born outside the current area of star forming regions, are more common than previously considered. 
\end{abstract}


\section{Introduction} \label{sec:intro}

Episodic or quasi-periodic dimming in the light curve of young stellar objects (YSOs), known as ``dippers,'',  have been discovered in the data obtained by recent satellite missions, {\it CoRoT}, {\it Spitzer} \citep{2010A&A...519A..88A,2011ApJ...733...50M,2014AJ....147...82C}, $K2$ \citep{2016ApJ...816...69A,2018MNRAS.476.2968H,2019MNRAS.483.3579A}, and {\it TESS} \citep{2019MNRAS.488.4465G}. \cite{2011ApJ...733...50M} revealed that dips are deeper in the optical wavelength than in the infrared bands, indicating that dips are shadows of circumstellar materials. However, the detailed mechanism has so far been poorly understood. Various mechanisms have been proposed such as accretion inflow onto a star \citep{1999A&A...349..619B, 2015A&A...577A..11M}, non-axisymmetric structure of the disk induced by the Rossby wave instability \citep{2015AJ....149..130S}, and transiting circumstellar clumps \citep{2016ApJ...816...69A}. 

Historically, new dippers are discovered through visual inspection-based approaches because their transit shape varies from object to object and changes over time. The first systematic survey using machine learning was done by \citep{2018MNRAS.476.2968H}. They searched for the dippers from $K2$ data and discovered about 100 dippers in famous star forming regions, Upper Scorpius and $\rho$ Ophiuchus.

While most dippers that have been found are located in the star forming regions and have the properties of M or K -type stars \citep{2015AJ....149..130S,2016MNRAS.462L.101A}, a few stars near the zero-age main sequence with episodic dimming have been also reported. RZ Psc was regarded as a variable star with IR excess in this stage \citep[30--50 Myr][]{2010A&A...524A...8G,2013A&A...553L...1D, 2018AJ....155...33P}. However, \citet{2019A&A...630A..64P} have identified the membership of RZ Psc with Cassiopeia--Taurus OB association from their proper motions and age as $20_{-5}^{+3}$ Myr updated by {\it Gaia} DR2 data \citep{2018A&A...616A...1G}. Recently, \cite{2019MNRAS.488.4465G} discovered episodic dimming in HD 240779, outside star forming regions in the {\it TESS} short cadence data. HD 240779 is between UX Ori stars and M-type dipper and belongs in the AB Doradus moving group. HD 240779 shows IR excess, but the moving group suggests that its age is $\approx125$Myr. At this age, primordial disks are thought to have already  dissipated, so this IR excess is thought to be due to a secondary disk. Though the origin of the disk is not fully understood, this discovery suggests that dimming events can be found even outside star forming regions. In this paper, we adopt the definition of dipper as a star with episodic or quasi periodic dimming with IR excess regardless of its age. In this sense, HD 240779 can be called an old dipper or debris dipper.  

YSO surveys have been done in star forming regions, but YSOs ($\lesssim 10$\,Myr) can also be found outside of star forming regions.
\citet{2019arXiv190807550M} found several runaway stars with disks around the Orion Nebula Cluster (ONC) using the proper motions obtained by {\it Gaia}. 
Runaway stars \citep{1954ApJ...119..625B,1986ApJS...61..419G} are stars with a peculiar velocity ($> 30 \kms$) that is higher than the velocity dispersion of young disk stars \citep{2014ApJ...793...51S}. Runaway stars initially found among OB-type stars, so called OB runaways \citep{1986ApJS...61..419G}. There are two mechanisms to form runaways: supernova explosion of a companion star in a tight binary \citep{1961BAN....15..265B} or binary-single stellar encounters in star clusters \citep{1986ApJS...61..419G}. Recent observations \citep{2010ApJ...715L..74E,2011MNRAS.416..501R,2013MNRAS.430L..20G,2018A&A...619A..78L,2019arXiv190807550M} and simulations \citep{2011Sci...334.1380F, 2012ApJ...746...15B,2012ApJ...753...85F} have suggested that young star clusters can actually produce runaway stars. 
Most recent study done by \citet{2019arXiv190807550M} reported that $\approx30\%$ of runaway young stars from the ONC have a disk. This suggests that protoplanetary disks can survive close stellar encounters, which are necessary for the formation of runaway stars. Therefore, we do not need to restrict the dipper survey to star forming regions. In this paper, we searched for dippers both inside and outside the star forming regions by systematically analyzing the full-frame images of {\it TESS} \citep{2015JATIS...1a4003R}.

The rest of the paper is organized as follows.In Section 2 we describe the pipeline to identify dippers from {\it TESS} full frame images. In Section 3, we show the statistical properties of 35 dippers we find. We also identify the memberships of these dippers. In section 4, We focus on the runaway dipper TIC 43488669. We discuss the origins of dippers and the occurrence of debris dippers in Section 5.

\section{Identification of Dippers}

\subsection{Aperture photometry of the {\it TESS} full-frame images}
In the survey, we performed the simple aperture photometry of the stars in the Candidate Target List version 8 \citep[CTLv8;][]{2019AJ....158..138S} using the {\it TESS} full-frame images \citep[FFIs;][]{2015ApJ...809...77S} of Sectors 1--13, corresponding to the first 1 year of the {\it TESS} mission. We extracted data cubes of a time-series of \( 13\times13\) pixels around the position of the stars in the CTLv8 from the FFIs. The aperture is defined by the pixels whose flux is 3 $\sigma$ times larger than the median flux of the 13 $\times$ 13 pixels and is connected to the central pixel by other pixels in the aperture. We also defined the 30 or more background pixels distinguished from its aperture whose flux does not exceed 0.5 
quartile deviation. We assume the median value of the background pixels as the background flux, which is constant for all of the target pixels. We subtracted the background flux from the light curve. We automatically removed poor quality data points caused by ``momentum dump'' and so on using time-series data of pointing of spacecraft provided in the FFIs. In this way, we created about $4 \times 10^6$ light curves in 30 minute cadence from FFIs.

\subsection{Convolutional Neural Network}

\begin{figure}
    \includegraphics[width=\linewidth]{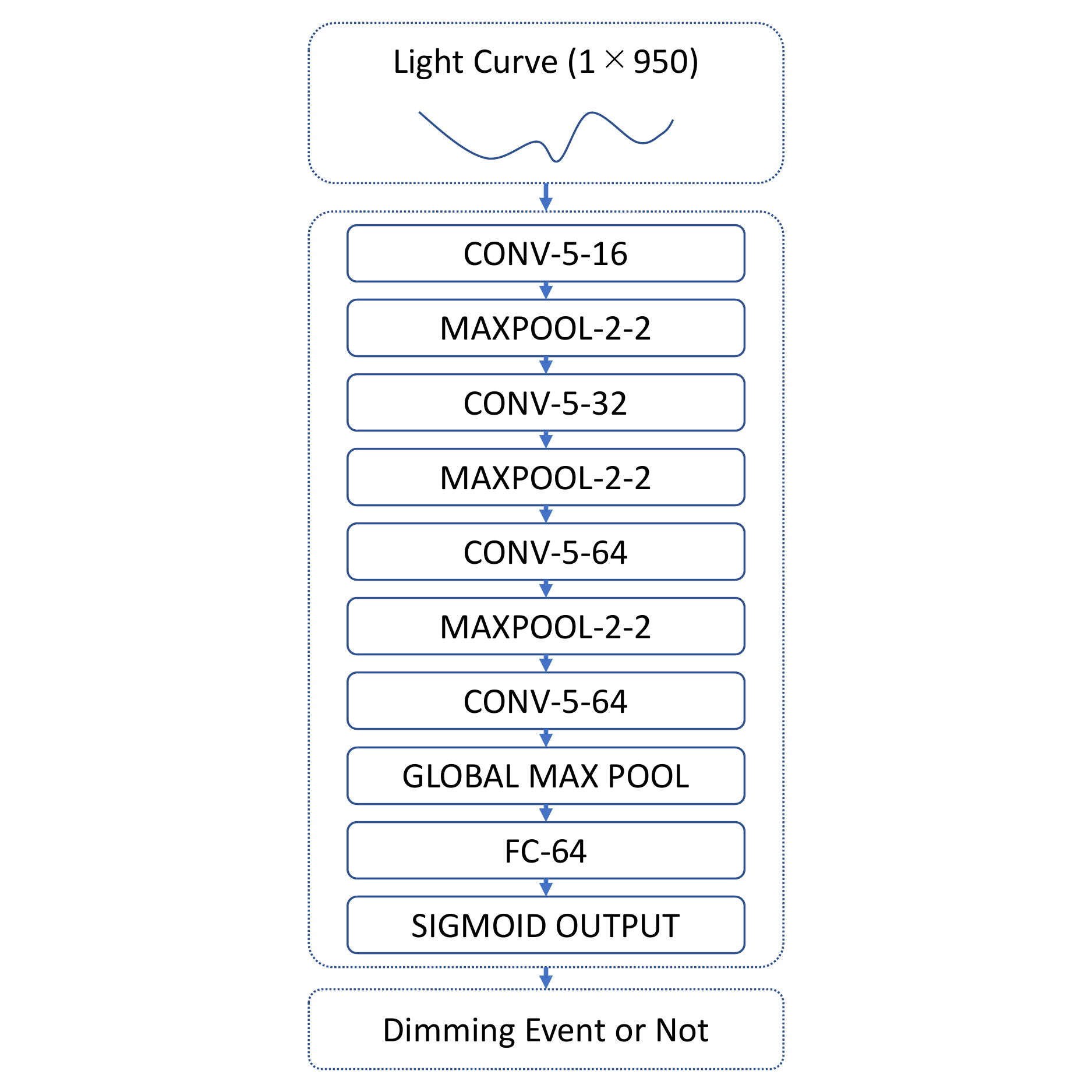}
    \caption{The convolutional neural network architecture used in this work. One dimensional time-series data of light curves are fed into four convolutional layers and one fully connected layer and ends in a sigmoid function. Following \citet{2018ApJ...869L...7A}, the convolutional layers are denoted as CONV-$\langle \mathrm{kernel \,size}\rangle$-$\langle \mathrm{number \,of \,feature \,maps}\rangle$, the max pooling layers are denoted as MAXPOOL-$\langle \mathrm{window \,length}\rangle$-$\langle \mathrm{stride \, length}\rangle$, and the fully connected layer is denoted as FC-$\langle \mathrm{number \,of \,units}\rangle$, where $\langle \mathrm{kernel \,size}\rangle$ indicates the length of the 1D convolution window, $\langle \mathrm{number \,of \,feature \,maps}\rangle$ is the dimensionality of the output space, and $\langle \mathrm{stride \, length}\rangle$ is the length of each window and $\langle \mathrm{number \,of \,units}\rangle$ means the dimensionality of the output space. The value of null in each light curve was removed and data were cut down to 950 points.} \label{fig:NNmodel}
\end{figure}

To find episodic events of the dippers from more than $4 \times 10^6$ light curves, we first identified objects that exhibited a large photometric variation such as eclipsing binaries by using a convolutional neural network (CNN). The CNN model is shown in Figure \ref{fig:NNmodel}, which is based on the Exonet-XS model \citep{2018ApJ...869L...7A}. We slightly modified the Exonet-XS model for {\it TESS} light curves; we do not use the local view (a light curve localized around a transit signal) used in the original Exonet-XS , because we do not have a common transit shape and we also do not use stellar parameters. We add one more convolutional layer to the Exonet-XS model. 

So far, we do not have sufficient training sets of dippers because HD 240779 is the only dipper found in {\it TESS}. Instead, we decided to use eclipsing binaries found in the {\it TESS} data as the training data set. This is because a large and sharp dip in the light curve of an eclipsing binary resembles those of dippers in a local view around a dip. 
First, we searched for eclipsing binaries from {\it TESS} light curves by finding outliers defined by the four quartile deviation, whose parameters such as threshold values was adjusted by Kepler Eclipsing Binary Catalog \citep{2016AJ....151..101A}. We detect light curves which have data points far from the baseline. A catalog of hundreds of {\it TESS} eclipsing binaries was created by visual inspection from the sets of light curves obtained by this process.

Then using these data, we trained a support vector machine, visually examined the data judged to be true, and found 4000 eclipsing binaries. We used {\tt thundersvm \citep{wenthundersvm18}} for the support vector analysis. Using these training data set, we trained our neural network model. For each phase, training sets for a negative detection were randomly chosen from the light curves. The final CNN identified $36,674$ eclipsing binary candidates. 

\subsection{Identifining and Screening of the Dipper Candidates}

\subsubsection{Visual Inspection}
We visually inspected all of the $36,674$ eclipsing binary candidates to classify the types of variation. We classified them into six types, stellar eclipse, stellar pulsation, variation by star spots, asteroid passing, systematic noise, and dippers. The eclipse by a companion or planet is completely cyclic and has the same shape in every transits. Although there are many types of the pulsating stars whose period is about from a day to a week such as Cepheid, RR Lyrae, Delta Scuti, and Gamma Doradus, all of them are almost periodic and well modeled as in the case of eclipse and transit signals. The variation by the star spots is also periodic, but it exhibits various shapes in the light curves. While the variation of the star spots, whose periodic shape can be roughly modeled by a smooth sine-like curves, is indistinguishable from the variation of the eclipsing binary, it can be easily distinguished from the dippers, whose shape exhibits complex structure. Also, a large single dip is observed when an asteroid crosses over the target star. Although the episodic dimming event owing to asteroid crossing is similar to the dipper event, it can be distinguished by examining pixels outside the aperture; if an asteroid crosses, pixels outside the aperture will also exhibits large variation. A systematic noise also induces large variations outside the aperture. We can distinguish the systematic noise from the dipper because the common variation pattern is observed in the other targets in the CCD chip. As a result, we discovered 41 dipper candidates in the {\it TESS} first one year data.

\subsubsection{Pixel-level Difference Imaging}

\begin{figure}
    \includegraphics[width=\linewidth]{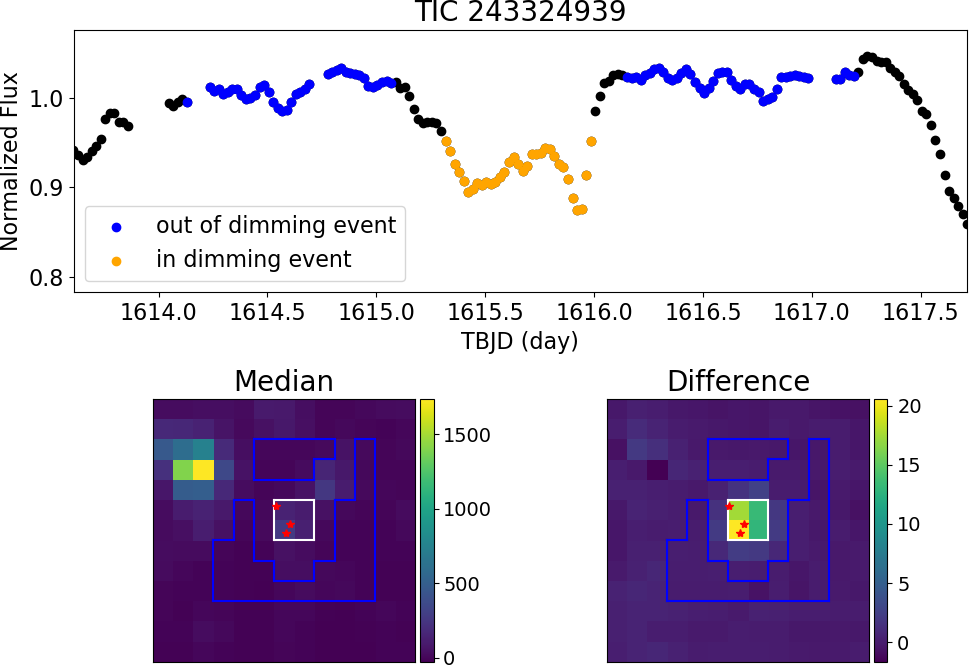}
    \caption{The pixel-level difference image of TIC 243324939. Black, blue and orange points in the upper panel are normalized flux of TIC 243324939. The red points in bottom images indicate stars, TIC 243324946, 243324939 and 243324936 from the top to the bottom. White range is an aperture for TIC 243324939 and blue regions are used to determine background flux. The bottom right image indicates that the dip occurred in the central four pixels around TIC 243324939. \label{fig:pixel_difference}}
\end{figure}

The contamination from nearby stars was investigated using the pixel-level difference imaging \citep{2013PASP..125..889B,2019AJ....157..218K}. Figure \ref{fig:pixel_difference} shows an example of the pixel-level difference image for TIC 243324939. We take the averages of pixels in and out of the dip (orange and blue) and compute the difference image (bottom right panel). The red points indicate stars, TIC 243324946, 243324939 and 243324936 from the top to the bottom, whose magnitudes in {\it TESS} band are 15.9, 13.7 and 13.6. We do not plot stars whose magnitudes are more than 16 in the figure. For this example, though both TIC 243324939 and 243324936 are located in the center pixel and have similar magnitudes, the difference image of the dip reveals that pixels around TIC 243324939 show large values; this implies that the source star that shows dimming event is TIC 243324939.

We also analyze the relative positions of the moment-weighted centroid of the dipper candidates using {\tt photutils}. All of their centroid shifts are consistent with our identification of the dimming stars.

Through the above process, we determined 37 dipping stars from contaminated images. 
However, we were not able to associate the dipping signal with any specific star for four systems by the above process.

\subsubsection{SED Analysis}
Young stellar objects generally exhibit the infrared excess owing to the circumstellar structure. We checked spectral energy distributions (SEDs) of the dipper candidates using the data of three bands ({\it $G_{BP}$, G, $G_{RP}$}) provided by {\it Gaia} DR2, three bands ({\it J, H, $K_{s}$}) from 2MASS, and four bands ($W1$, $W2$, $W3$, $W4$) from {\it WISE}. Each SED was compared with the best-fit black body model obtained by MCMC using {\it $G_{BP}$, G, $G_{RP}$, J, H, $K_{s}$} bands. The distance from {\it Gaia} is fixed, and the effective temperature and radius are regarded as free parameters. We compared observed {\it WISE} fluxes with the expected energy distributions derived by black body model to check the infrared excess. Comparing with the result of Virtual Observatory SED Analyzer \citep[VOSA][]{2008A&A...492..277B}, we excluded one target with no clear infrared excess from our list of dippers because we focus on the clear examples of the dippers with infrared excess in this paper.

\subsubsection{Other Variables}
There are variable stars which shows infrared excess other than T-Tauri stars. For instance, asymptotic giant branch (AGB) stars and R Coronae Borealis (RCB) stars exhibit such excess in their SEDs. Both have infrared excess due to circumstellar envelopes and sometimes exhibit quasi-periodic or aperiodic variability. Because both AGBs and RCBs are giant stars, we used the stellar radius provided by {\it TESS} Input Catalog version 8 \citep[TICv8;][]{2019AJ....158..138S} to exclude such stars from dipper candidates. We dropped one object with a stellar radius greater than $3 R_\odot$. We finally identified 35 dippers as listed in Table \ref{tab:stellar_properties}.
We additionally analyzed Reduced Unit Weighted Error to distinguish multiple stars from each other, as summarized in Appendix.




\section{Disk and stellar properties of dippers}\label{sec:result}
In this survey, we discovered 35 dippers. All light curves are shown in Figure \ref{fig:lc1}. All of them exhibit infrared excess in their SED diagrams. There are a wide variety of the shape, duration, and frequency of the dips. The depths of dips are typically from 10 percent to several ten percent. Stellar properties of the dippers from {\it Gaia} DR2 and TICv8 are listed in Table \ref{tab:stellar_properties}.


\begin{table*}
\caption{Properties of dippers \label{tab:stellar_properties}}
\begin{center}
\begin{tabular}{cccccccccc} 
 \hline
 \hline
 TIC ID\footnotemark[1] & Gaia\footnotemark[2] & Membership\footnotemark[3] & Disk Type\footnotemark[4] & $l$\footnotemark[5] & $b$\footnotemark[5] & parallax\footnotemark[5] & Tmag\footnotemark[6] & Teff\footnotemark[6] & Radius\footnotemark[6]\\
 &  &  &  & deg & deg & mas &  & K & $R_{\odot}$\\
 \hline
24775707 & 3208869061545298048 & Orion & T? & 209.191 & -21.826 & $3.02\pm0.03$ & 12.9 & $4223\pm129$ & $1.23\pm0.12$\\
34397579 & 3208985468042433792 & Orion & F & 208.992 & -21.071 & $2.89\pm0.05$ & 9.88 & $6330\pm140$ & $2.3\pm0.12$\\
37947642 & 3179285734830216192 &  & E & 206.123 & -37.112 & $3.99\pm0.18$ & 15.52 & $3082\pm157$ & $0.48\pm0.02$\\
38044097 & 3179048313334809472 &  & E & 206.93 & -36.098 & $4.69\pm0.07$ & 14.62 & $3210\pm157$ & $0.55\pm0.02$\\
43488669 & 3183786280737192704 &  & E & 207.954 & -28.688 & $3.87\pm0.02$ & 12.73 & $5193\pm117$ & $0.64\pm0.04$\\
47390297 & 6607586870253814144 &  & F & 19.839 & -62.749 & $1.48\pm0.05$ & 11.45 & $5997\pm133$ & $2.29\pm0.14$\\
50745680 & 3216722873103020288 &  & F & 206.566 & -19.14 & $3.5\pm0.21$ & 15.4 & $3190\pm161$ & $0.63\pm0.03$\\
57830249 & 3478940625208241920 & TW Hydrae & E & 285.143 & 29.71 & $20.52\pm0.1$ & 11.96 & $3085\pm157$ & $0.49\pm0.01$\\
67822046 & 6041816673812516864 & ScoOB2 & F & 347.056 & 16.905 & $6.57\pm0.04$ & 12.49 & $3662\pm122$ & $0.98\pm0.12$\\
81198421 & 5519376920648594304 & Vel OB2 & F & 262.533 & -7.188 & $2.84\pm0.02$ & 12.66 & $4655\pm127$ & $1.3\pm0.09$\\
124072515 & 5530952441620701440 & Vel OB2 & F & 259.267 & -10.03 & $2.56\pm0.02$ & 12.99 & $4381\pm126$ & $1.33\pm0.11$\\
144697607 & 3336447113703238784 & Orion & F & 195.72 & -11.048 & $2.54\pm0.1$ & 15.3 & $3290\pm159$ & $0.71\pm0.03$\\
167303776 & 5280226578890455168 & Carina & E & 278.41 & -26.817 & $10.67\pm0.06$ & 14.15 & $3149\pm157$ & $0.35\pm0.01$\\
226241509 & 5944710143324257792 & ScoOB2 & F & 340.049 & 4.136 & $6.67\pm0.05$ & 11.57 & $3954\pm127$ & $1.32\pm0.14$\\
243324939 & 6083123916915057664 & ScoOB2 & E & 310.949 & 14.543 & $7.93\pm0.09$ & 13.61 & $3285\pm158$ & $0.54\pm0.02$\\
248473576 & 3213713956452945024 & Orion & E/T & 204.159 & -23.562 & $2.66\pm0.12$ & 15.6 & $3214\pm157$ & $0.61\pm0.03$\\
248992945 & 3215349651798096768 & Orion & T? & 202.128 & -22.341 & $2.92\pm0.03$ & 12.31 & $4564\pm126$ & $1.39\pm0.11$\\
266079454 & 5336376503118529536 & ScoOB2 & E & 293.79 & 2.527 & $9.75\pm0.08$ & 14.22 & $3030\pm157$ & $0.4\pm0.01$\\
276070646 & 6107552866260657408 & ScoOB2 & E & 314.213 & 15.951 & $6.65\pm0.16$ & 14.26 & $3196\pm157$ & $0.5\pm0.02$\\
276664304 & 3017275483925567872 & Orion & F & 209.697 & -19.194 & $2.55\pm0.03$ & 12.97 & $6158\pm180$ & $1.21\pm0.07$\\
284730577 & 3291818722708659200 & AB Doradus & E & 190.329 & -17.322 & $10.53\pm0.06$ & 9.25 & $5677\pm131$ & $0.98\pm0.05$\\
317873721 & 2995720039486652672 &  & F & 219.838 & -18.11 & $1.51\pm0.04$ & 11.94 & $8211\pm148$ & $1.67\pm0.06$\\
323354855 & 5209023133585929984 & ScoOB2 & F/E & 292.586 & -21.437 & $10.14\pm0.03$ & 10.1 & $4308\pm150$ & $1.22\pm0.13$\\
334999132 & 5843724192198636416 & ScoOB2 & E & 303.399 & -8.46 & $9.72\pm0.05$ & 14.02 & $3018\pm157$ & $0.46\pm0.01$\\
361957000 & 5789045856890265216 & ScoOB2 & F & 303.692 & -13.981 & $5\pm0.03$ & 11.49 & $4064\pm129$ & \\
412308868 & 6061152101197951104 & ScoOB2 & E & 301.848 & 5.8 & $8.75\pm0.05$ & 12.71 & $3496\pm118$ & $0.74\pm0.1$\\
419954392 & 6069015327310399104 & ScoOB2 & E & 309.034 & 10.088 & $7.29\pm0.13$ & 14.8 & $3230\pm159$ & $0.45\pm0.01$\\
427334205 & 3209524233037903360 & Orion & F & 208.721 & -19.567 & $2.43\pm0.03$ & 12.71 & $4777\pm69$ & \\
427395967 & 3017187110667027200 & Orion & F & 209.595 & -19.513 & $2.63\pm0.06$ & 15.37 & $5321\pm319$ & \\
427399794 & 3217577258060423808 &  & F & 205.343 & -17.396 & $4.66\pm0.53$ & 14.3 & $3473\pm158$ & $0.6\pm0.05$\\
434229695 & 3304172564776041728 &  & F & 181.268 & -27.171 & $3.4\pm0.04$ & 11.34 & $5312\pm147$ & $1.92\pm0.11$\\
442549866 & 6011594294619304064 & ScoOB2 & E & 338.978 & 15.028 & $9.55\pm0.09$ & 14.79 & $3114\pm160$ & $0.36\pm0.01$\\
452694529 & 5218060397614069120 & ScoOB2 & E & 290.321 & -15.185 & $7.39\pm0.03$ & 12.73 & $3572\pm115$ & $0.83\pm0.11$\\
454365494 & 5201143758382913792 & Chamaeleon? & E & 297.654 & -15.686 & $5.29\pm0.08$ & 14.39 & $3101\pm124$ & \\
457231768 & 3221803265361341056 & Orion & F & 202.573 & -18.473 & $2.19\pm0.08$ & 9.88 & $8785\pm158$ & $1.96\pm0.09$\\
    \hline
\end{tabular}
\end{center}
\footnotetext[1]{{\it TESS} Input Catalog ID}
\footnotetext[2]{{\it Gaia} Source ID associated with TIC ID}
\footnotetext[3]{The possible membership determined in Section \ref{sec:result}. The blank means that we cannot identify the membership.}
\footnotetext[4]{Disk Type determined from Figure.\ref{fig:disk}: F=full, E=evolved, T=transitional. Dippers with two classes are those whose classes we cannot identify with Figure.\ref{fig:disk}}
\footnotetext[5]{Galactic longitude, latitude and parallax listed in {\it Gaia} DR2}
\footnotetext[6]{{\it TESS} magnitude, effective temperature, and stellar radius listed in TICv8}
\end{table*}


Although all of the dippers we found showed IR excess possibly resulting from disks, the extent of the excess is different, which means that each dipper has different disk structure. We classified the dippers according to their IR colors. \citet{2012ApJ...758...31L} classified the disk structure in the members of Upper Scorpius (USco) with the color-color diagram. They used {\it $K_{s}$} band, which is dominated by emission from stellar photosphere, and $W3$ and $W4$ bands as emission from a disk. They corrected the extinction of {\it $K_{s}$}, $W3$ and $W4$ bands. Then, they subtracted the component of IR colors from stellar photospheres assuming the same spectral type stars. The IR colors calculated by this process are denoted by $E(K_{s}-W3), E(K_{s}-W4)$. 

\begin{figure}
    \begin{center}
        \includegraphics[width=\linewidth]{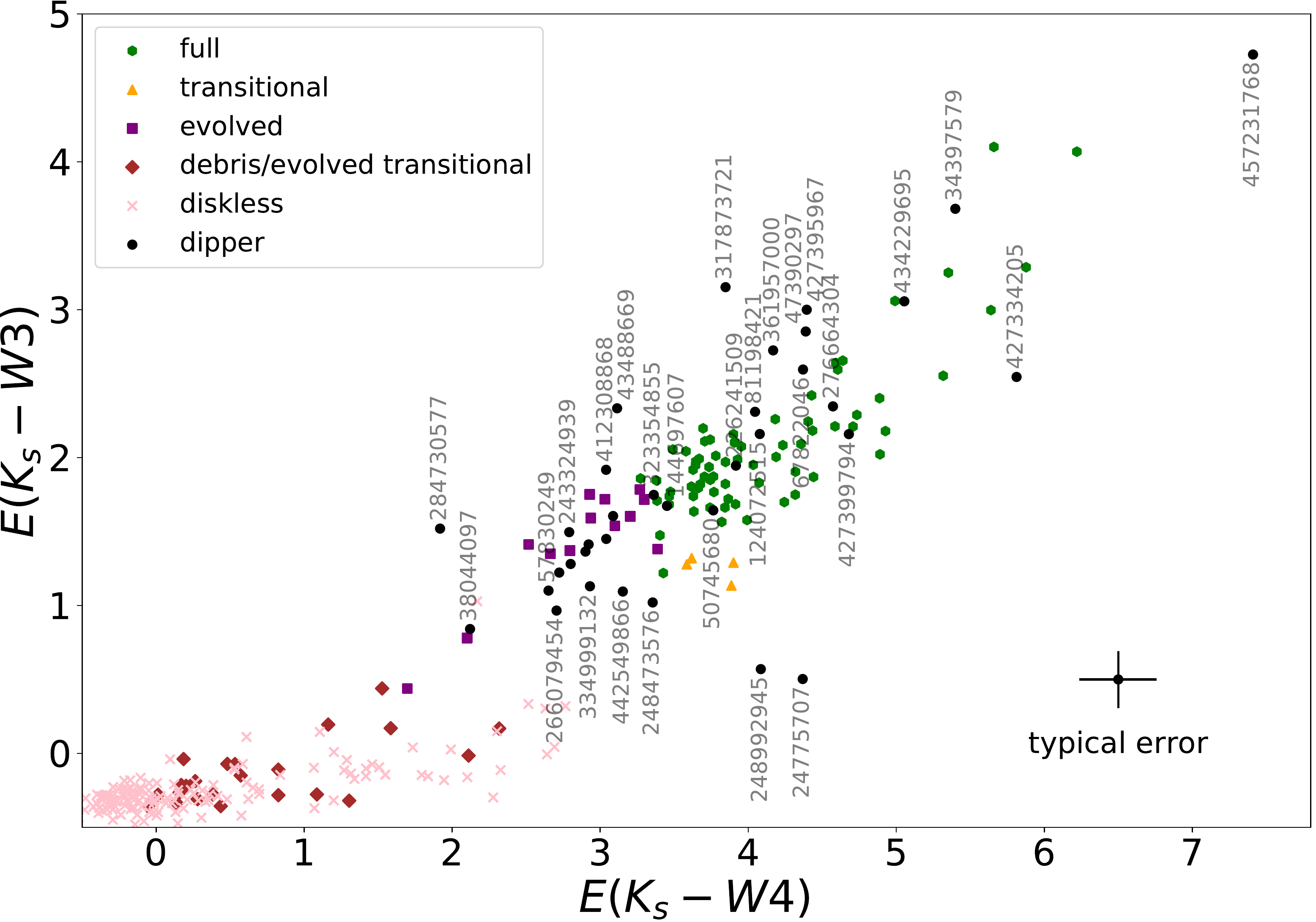}
        \caption{IR color excess for classification of the disk type. The $x$ and $y$ axes are the extinction-corrected IR excess $(K_{s} - W4)$ and $(K_{s} - W3)$ relative to the stellar photosphere. The colored markers are the members of Upper Sco with different disk types listed in \citet{2012ApJ...758...31L}. The black points with labels indicate our dippers. We do not put labels for TIC
        167303776, 276070646, 419954392, 37947642 and 454365494 in the crowded region, but all of them are classified as an evolved system. The typical error embedded in the panels is estimated with the typical errors of temperature in TIC and magnitudes of 2MASS and {\it WISE}. \label{fig:disk}}
    \end{center}
\end{figure}

Using 2MASS and {\it WISE} data, we re-computed their $E(K_{s}-W3), E(K_{s}-W4)$ values in their catalog and compare them with those of our dippers. The extinction was corrected as follows: we first obtained $A_{Ks}$ values from the dust map {\tt mwdust} \citep{2016ApJ...818..130B}; then we calculated $A_{W3}$ and $A_{W4}$ according to \citet{2016ApJS..224...23X}. 
Figure \ref{fig:disk} shows the classification of the disk stages. The colored markers represent USco members with different disk types and the black circles indicate our dippers. We estimated a typical error of $E(K_{s}-W3)$ and $ E(K_{s}-W4)$ by propagating typical errors of the temperature given in the TIC and magnitude of 2MASS and {\it WISE}. Most of the dippers are classified as a full or evolved system. As exceptions, TIC 248992945 and 24775707 appear not to belong any disk classes drawn in the diagram. Because these two dippers exhibit a weak emission in $W3$ band and a strong emission in $W4$ band, which implies that they have a large inner cavity, we classified them as a transitional system. No dipper is classified as a debris system. We summarize the classes in Table \ref{tab:stellar_properties}.

\begin{figure*}
 \begin{center}
     \includegraphics[width=0.49\linewidth]{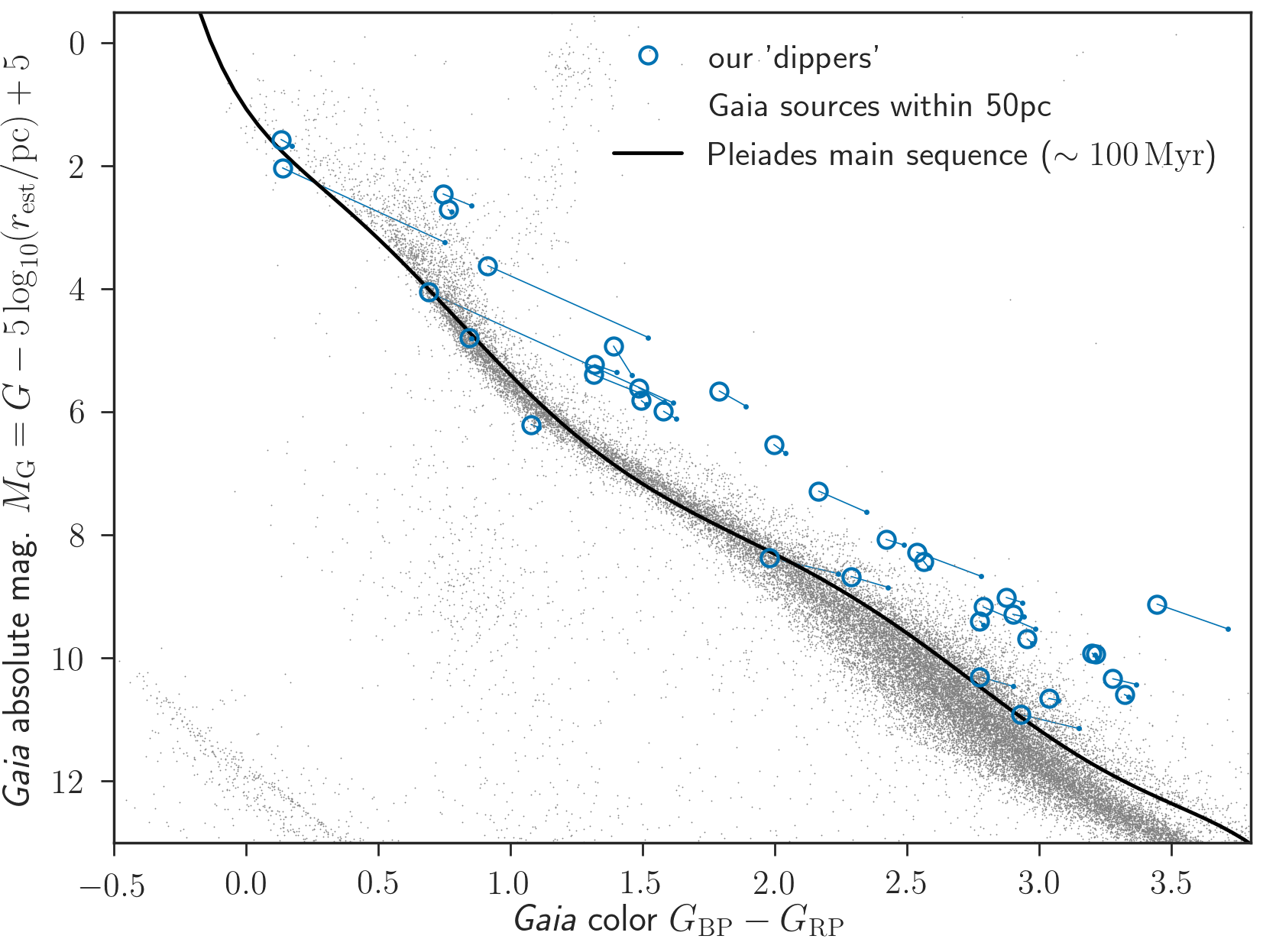}
     \includegraphics[width=0.49\linewidth]{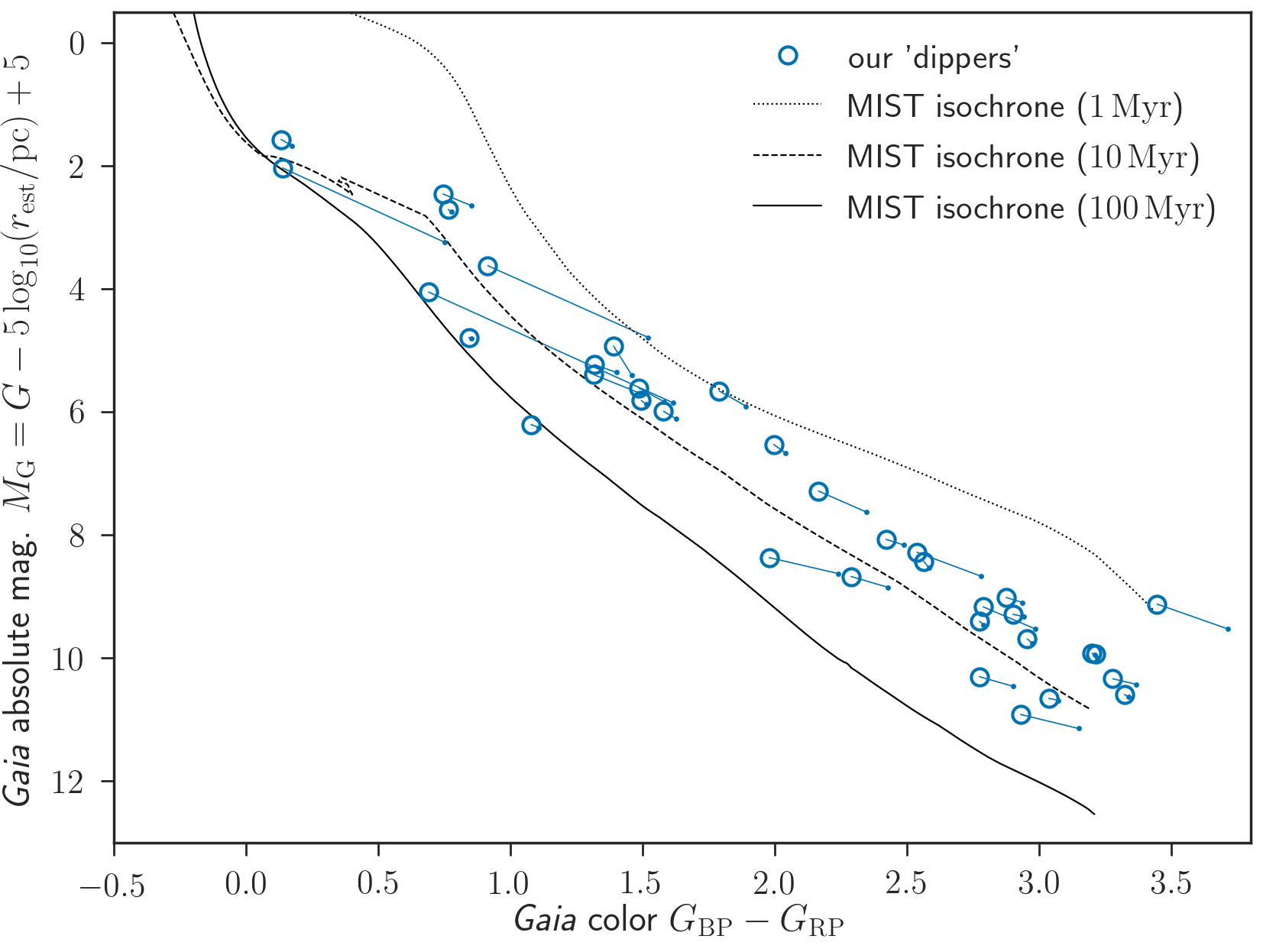}
    \caption{The {\it TESS} dippers on the HR diagram. The open circles show the absolute magnitudes in the {\it Gaia} $G$-band against {\it Gaia} colors both corrected for extinction, and they are connected to the values without correction (smaller dots) with solid lines; see Section \ref{sec:result} for details. {\it Left} --- Comparison with stars within $50\,\mathrm{pc}$ from the {\it Gaia} DR2 (gray dots), as well as the main sequence of the Pleiades cluster (black solid line). {\it Right} --- Comparison with MESA isochrones for the ages of 1~Myr (dotted line), 10~Myr (dashed line), and 100~Myr (solid line).\label{fig:HR}}
 \end{center}
 \end{figure*}

In Figure \ref{fig:HR}, we construct the Hertzsprung--Russell diagram of our dippers using the {\it Gaia} DR2 photometry and parallax, where the absolute magnitude in the {\it Gaia} $G$-band was calculated from the distance $r_{\rm est}$ estimated by \citet{2018AJ....156...58B}. 
The open circles show the values corrected for interstellar extinction, and they are connected with lines to smaller dots that show the values without the correction. 

We attempted to correct for the extinction as follows:
we first retrieved $E(B-V)$ from the dust map {\tt mwdust} \citep{2016ApJ...818..130B} and converted it to $A_V=3.1E(B-V)$; we then used the effective temperature of each star and the empirical relation in \citet{2018A&A...614A..19D} to convert $A_V$ into the total extinction in the $G$-band $A_G$; finally $A_G$ was converted to $E(G_{\rm BP}-G_{\rm RP})$ interpolating the results by \citet{2010A&A...523A..48J} and assuming $G_{\rm BP}-G_{\rm RP}\approx (V-I_C)_0$. 
In the left panel, our dippers are compared to nearby stars within $50\,\mathrm{pc}$ from {\it Gaia} DR2 (small gray dots), as well as the main sequence of the Pleiades cluster (solid line) constructed by \citet{2019AJ....158..190H}. In the right panel, the dippers are compared to the stellar isochrones \citep{2016ApJS..222....8D, 2016ApJ...823..102C} computed with Modules for Experiments in Stellar Astrophysics \citep[MESA;][]{2011ApJS..192....3P,2013ApJS..208....4P,2015ApJS..220...15P} for ages of $1$, $10$, and $100\,\mathrm{Myr}$ and for $\mathrm{[Fe/H]}=0$. 
Both of these comparisons suggest that our dippers, as a whole, have luminosities and colors consistent with low-mass stars younger than $\sim100\,\mathrm{Myr}$.
Here we do not attempt to draw any further quantitative conclusion because the accuracy of the dust map may be limited for those stars in star forming regions and because some of our dippers might exhibit strong circumstellar extinction that we did not take into account.
That said, the above qualitative conclusion would remain unchanged even if we underestimate $A_{\rm G}$ by $\sim1$ due, for example, to ignoring circumstellar extinction \citep{2016ApJ...816...69A}.


\begin{figure*}
\begin{center}
    \includegraphics[width=0.95\linewidth]{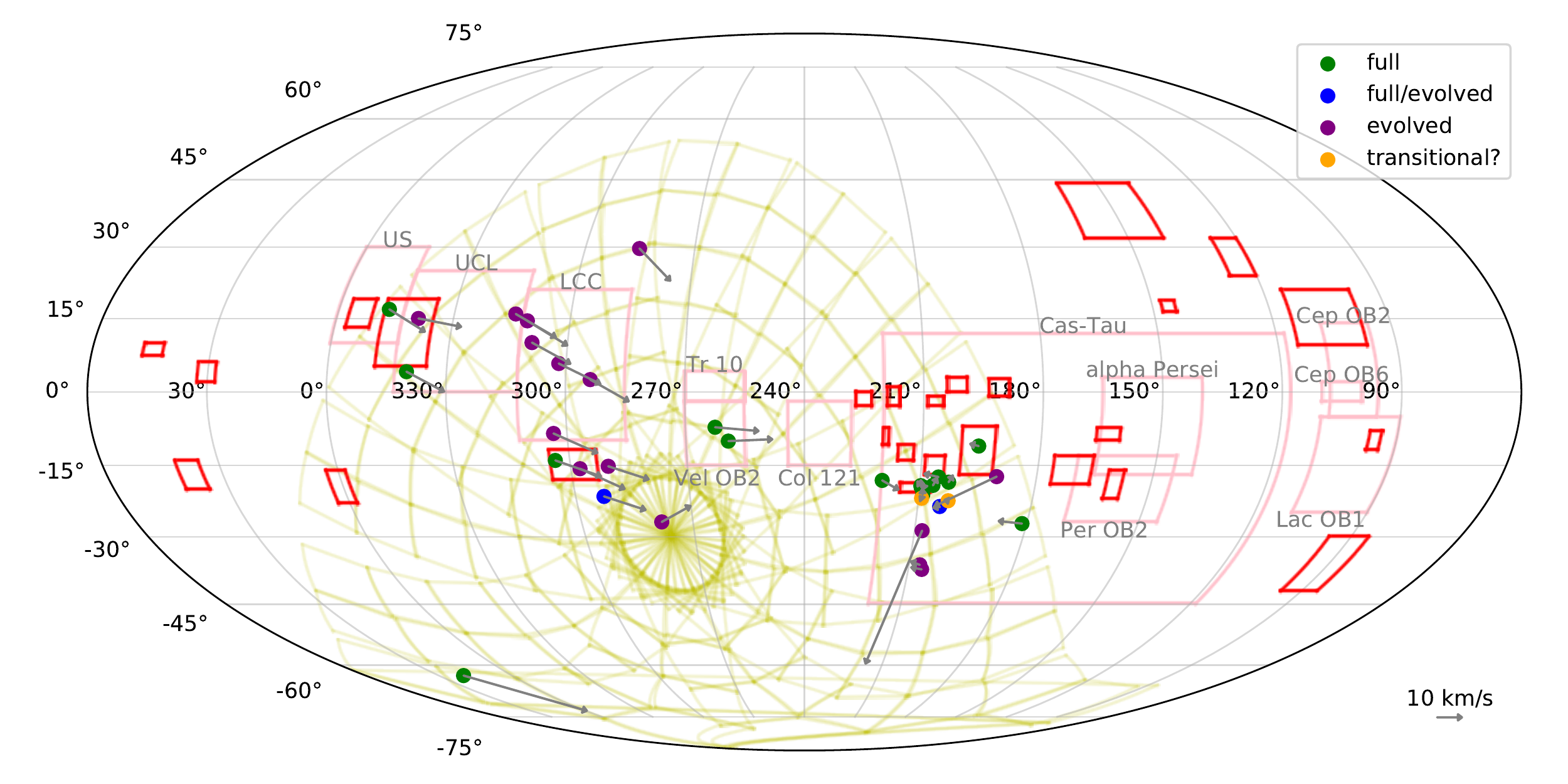}
    \caption{The distribution of the {\it TESS} dippers on the galactic coordinate. Each color of dipper represents disk type classified by Figure \ref{fig:disk}. The Evolved/Transitional dipper is colored by blue as well as well as Full/Evolved disk types. The red and pink rectangles show the approximate locations of nearby star forming regions and OB associations. The names of OB associations are labeled in this figure. The yellow area indicates the survey region in the first-year data of {\it TESS}. The gray arrows indicate tangential velocities of the dippers. The length of gray arrows is proportional to the tangential velocities ($\kms$) calculated from {\it Gaia} DR2 data. The arrow corresponding to 10\kms velocity is shown in at the bottom. \label{fig:galactic}}
\end{center}
\end{figure*}

Figure \ref{fig:galactic} shows the spatial distribution of the dippers in galactic coordinates. The survey area is shown by yellow rectangles. The markers with arrows indicate our dippers with proper motion, colored based on the classification of the disk type.  While most of the dippers are located in the star forming regions as indicated by red \citep[nearby molecular clouds;][]{2019ApJ...879..125Z} and pink \citep[nearby OB association;][]{1999AJ....117..354D} rectangles, some of them are far from any star forming regions. Moreover, some dippers have high proper motion as shown by gray arrows.

\section{Characterization of individual stars}
Next, we analyzed the individual dippers to determine their memberships and ages. We mainly used kinematic data of their positions, parallaxes, and proper motions obtained from {\it Gaia} DR2 data to determine their memberships. Dividing the dippers into several areas according to their positions (Sco OB2 and Chamaeleon, Vel OB2, Orion, and Field), we compare their kinematics with that of the stars that associates with known clusters. we additionally performed follow-up observations of four out of the six dippers near the Orion (TIC 434229695, 284730577, 317873721, and 43488669), and also TIC 34397579 and TIC 457231768 with high-dispersion spectra obtained by High-Dispersion Spectrograph \citep[HDS][]{2002PASJ...54..855N} installed on the Subaru telescope. The observation was performed on UT 16 September 2019 with the non-standard setup and a 2 $\times$ 1 binning without the image rotator. We use the image slicer (IS) \#2 with a spectral resolution of about 80,000. The non-standard setup of the wavelength range includes H$\alpha$ and lithium 6707 \AA\, line as an indicator of YSOs and also Mg triplet lines (5165--5185 \AA) for estimates of stellar temperature, log $g$, and radial velocity. The data were reduced by the standard procedure (the bias subtraction, flat fielding, order tracing/extraction, and wavelength calibration), which yields one-dimensional spectra with a signal-to-noise ratio of 80--100 per pixel at 6563 \AA.

\subsection{Sco OB2 and Chamaeleon}

\begin{figure}
    \begin{center}
        \includegraphics[width=\linewidth]{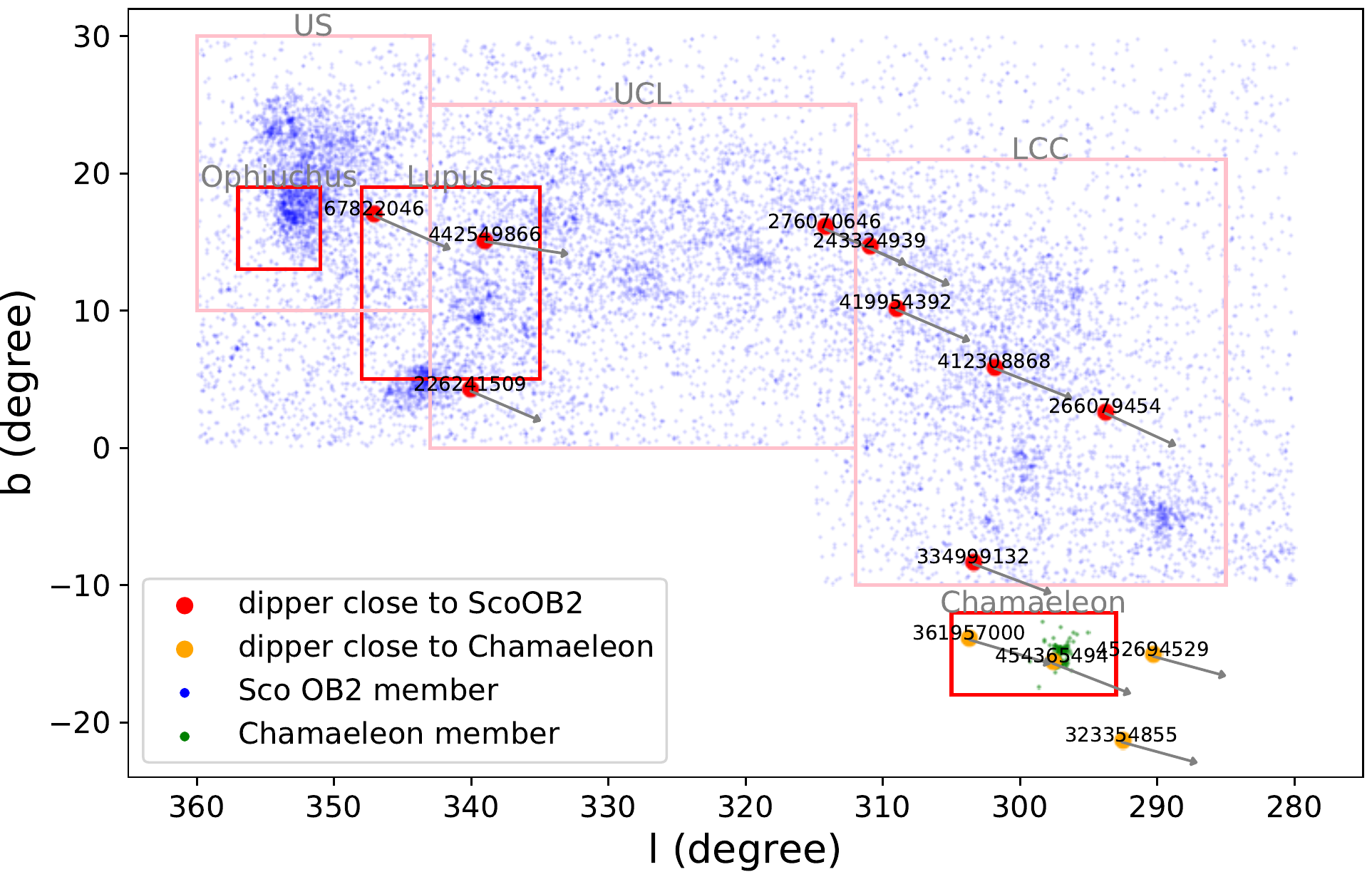}
        \caption{The galactic positions of Sco OB2 members, Cha I members, and the 13 dippers located near these systems. The blue and green dots indicate the members of Sco OB2 and Cha I listed in \citet{2019A&A...623A.112D} and \citet{2007ApJS..173..104L}. The red and orange points are dippers located close to Sco OB2 and Cha I in the celestial plane.\label{fig:Sco_plot}}
    \end{center}
\end{figure}

We found 13 dippers near the Scorpius-Centaurus OB Association (Sco OB2) and Chamaeleon complex (Chamaeleon). Sco OB2 is the nearest OB association to the Sun. Sco OB2 consists of three large sub-associations, Upper Scorpius (USco), Upper Cen-Lup (UCL),  and Lower Cen-Crux (LCC). Their mean ages were estimated by \citet{2012ApJ...746..154P}, $\sim 11$Myr old for US, which is the youngest part of Sco OB2, $\sim 16$Myr and $\sim 17$Myr for UCL and LCC. Thanks to {\it Gaia} mission, today, more than 10,000 pre-main sequence members for them have been discovered \citep{2019A&A...623A.112D}. The dippers, TIC 67822046, 226241509, 266079454, 276070646, 334999132, 412308868, 419954392, 442549866, and 243324939 are located close to Sco OB2 (shown in Figure\ref{fig:Sco_plot} as red circles). 

Chamaeleon is one of the nearest star forming regions and consists of three sub regions, Chamaeleon I (Cha I), Chamaeleon I\hspace{-.1em}I (Cha I\hspace{-1pt}I), and Chamaeleon I\hspace{-.1em}I\hspace{-.1em}I (Cha {I\hspace{-.1em}I\hspace{-.1em}I}) \footnote{Because of the lack of multi-band photometric data, \citet{2019ApJ...879..125Z} do not determine a precise area of Cha {I\hspace{-.1em}I\hspace{-.1em}I}. Therefore, we did not include Cha {I\hspace{-.1em}I\hspace{-.1em}I} in Figure \ref{fig:galactic}. } The catalog of Cha I members is provided by \citet{2007ApJS..173..104L}. They calculated the isochrone ages of southern and northern clusters of Cha I as 3--5 and 4--6 Myr. TIC 323354855, 361957000, 452694529 and 454365494 are located close to Chamaeleon (shown in Figure \ref{fig:Sco_plot} as orange circles). We note that we do not plot the members of Cha {I\hspace{-.1em}I} and Cha {I\hspace{-.1em}I\hspace{-.1em}I} because no sufficient information on the population has been reported yet.

Checking the population catalogs of Sco OB2 and Cha I by \citet{2019A&A...623A.112D} and \citet{2007ApJS..173..104L}, we find that TIC 67822046, 226241509, 266079454, 276070646, 334999132, 412308868 and 243324939 are known as the members of Sco OB2. The others are not listed in these catalogs.

\begin{figure}
    \begin{center}
        \includegraphics[width=\linewidth]{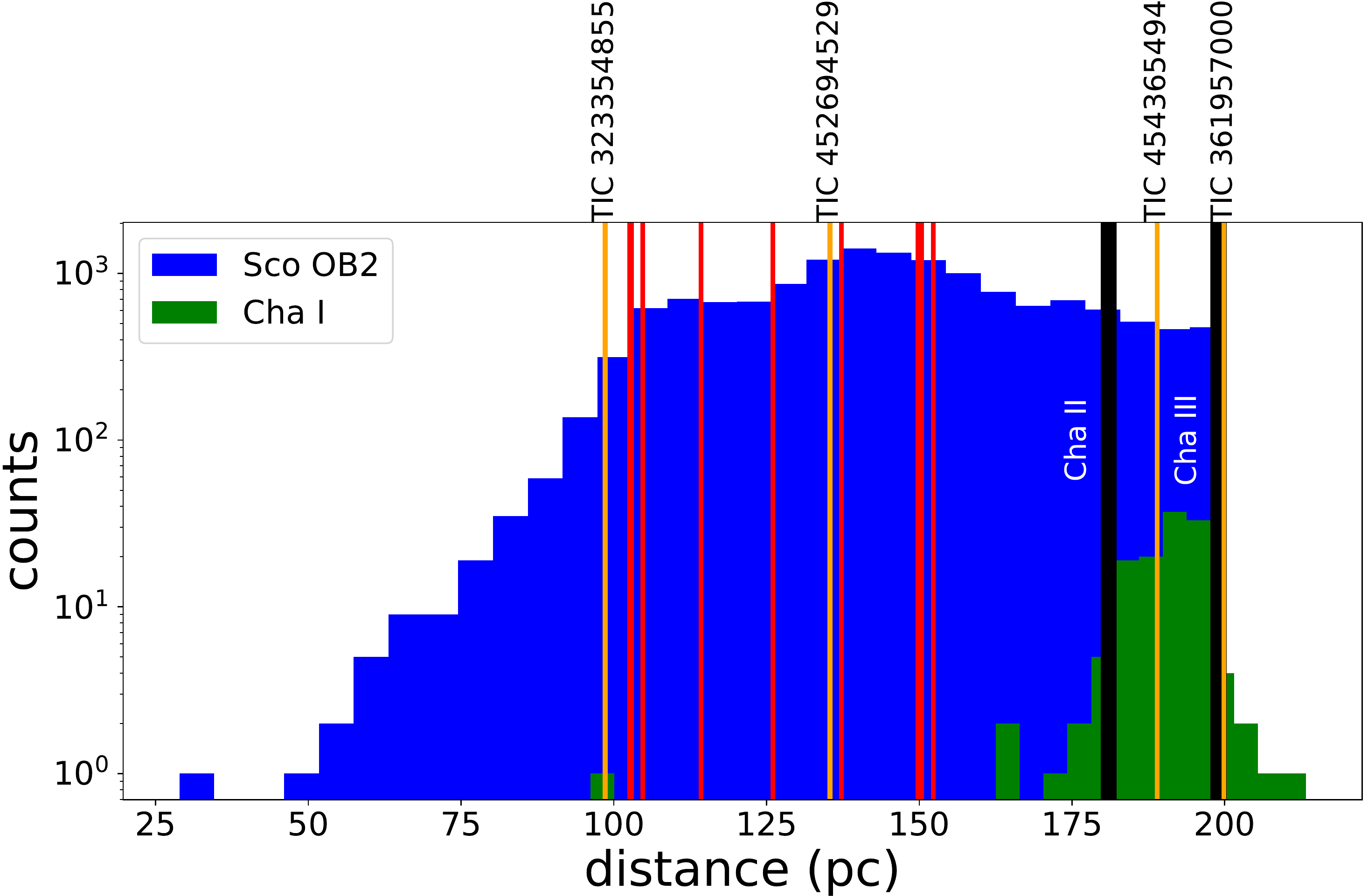}
        \caption{The distance distribution for members of Sco OB2 and Cha I. The red and orange lines represent the distance of dippers whose colors inherits Figure \ref{fig:Sco_plot}. We only show TIC IDs for dippers close to Cha I. The black vertical lines represent the average distances of Cha{I\hspace{-.1em}I} and Cha{I\hspace{-.1em}I\hspace{-.1em}I}. \label{fig:ScoOB2_distance}}
    \end{center}
\end{figure}

To identify the memberships of the other six dippers not listed in these catalogs, we analyzed their distances and kinematic properties using {\it Gaia} DR2 data. 
The fundamental astrometric parameters (positions, parallaxes, and proper motions) provided {\it Gaia} DR2 enabled us to well determine the memberships from the dippers' positions and kinematic natures. 
Figure \ref{fig:ScoOB2_distance} shows the distributions of the distances in the Sco OB2 (blue) and Cha I (green) catalogs. The dippers close to Sco OB2 and Chamaeleon in the celestial plane are indicated by the red and orange lines. Because \citet{2019A&A...623A.112D} surveyed only Sco OB2 members whose parallaxes are larger than 5, the outer steep edge of Sco OB2 shown in Figure \ref{fig:ScoOB2_distance} can be explained by the selection bias. Note that they have also searched for the members of Sco OB2 whose parallaxes are $3.3 < \pi < 5$ mas and found that no significant presence of Sco OB2 members beyond 200 pc. We also indicate the average distances of Cha {I\hspace{-.1em}I} and Cha {I\hspace{-.1em}I\hspace{-.1em}I} as the black lines. Their distances ($d_{ChaI\hspace{-.1em}I}$ and $d_{ChaI\hspace{-.1em}I\hspace{-.1em}I}$) are estimated by \citet{2018A&A...610A..64V} as $181^{+11}_{-10}$ pc and $199^{+12}_{-11}$ pc. From the distances, all the dippers close to Sco OB2, TIC 323354855 and 452694529 are consistent with the memberships of Sco OB2. TIC 361957000 and 454365494 can belong either to Sco OB2 or to Chamaeleon.

\begin{figure*}
    \begin{center}
        \includegraphics[width=\linewidth]{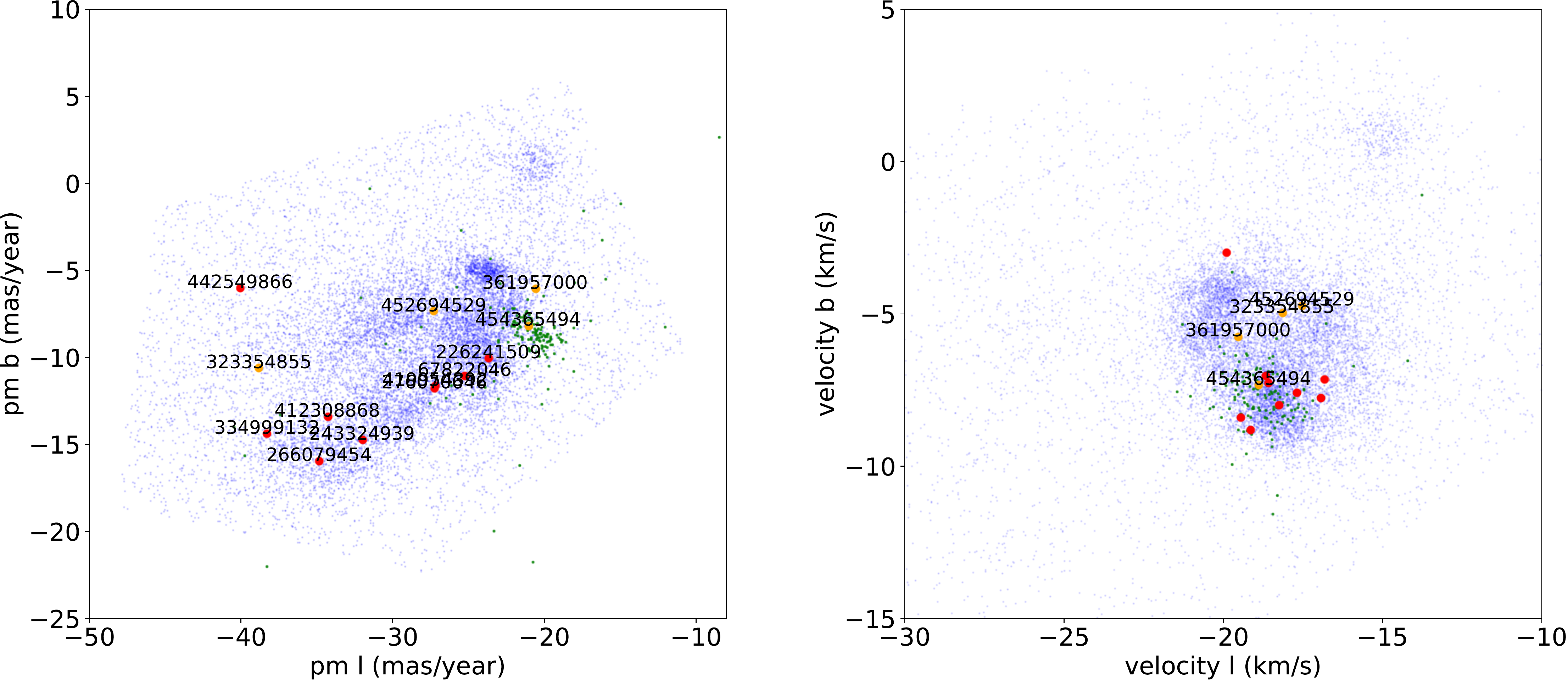}
        \caption{The distribution of proper motion and tangential velocities of Sco OB2 (blue), Cha I (green), and the 13 dippers (red and orange) located near these systems. The right and left panels have the different unit of the velocities corresponding to milliarcsecond $\mathrm{(mas)\, year^{-1}}$ and \kms. In the right panel, we only show TIC ID of dippers located close to Cha I in the celestial plane. The color of each star inherits Figure \ref{fig:Sco_plot}.} \label{fig:ScoOB2}
    \end{center}
\end{figure*}

Figure \ref{fig:ScoOB2} shows the distributions of the proper motions and tangential velocities in the Sco OB2 (blue) and Cha I (green) catalogs. The kinematics also shows that all the dippers close to Sco OB2 are consistent with the memberships of Sco OB2. TIC 452694529 and 361957000 are slightly separated from Cha I but well agrees with Sco OB2. Because TIC 323354855 and 454365494 are consistent with both clusters, we cannot determine their memberships only by the kinematics. Considering their positions, distances and kinematic nature, we conclude that TIC 454365494 is possibly a member of Chamaeleon and other dippers near the Sco OB2 and Chamaeleon are members of Sco OB2.


\subsection{Vel OB2}
Vel OB2 is the OB association and contains $\gamma^{2}$ Velorum, which is a spectroscopic binary composed of a massive O star and a Wolf-Rayet star\citep{2007MNRAS.377..415N}. Using the {\it Gaia} DR2 data, \citet{2018MNRAS.481L..11B} reported that Vel OB2 consisted of six kinematic clusters with the different ages.

\begin{figure}
    \begin{center}
        \includegraphics[width=\linewidth]{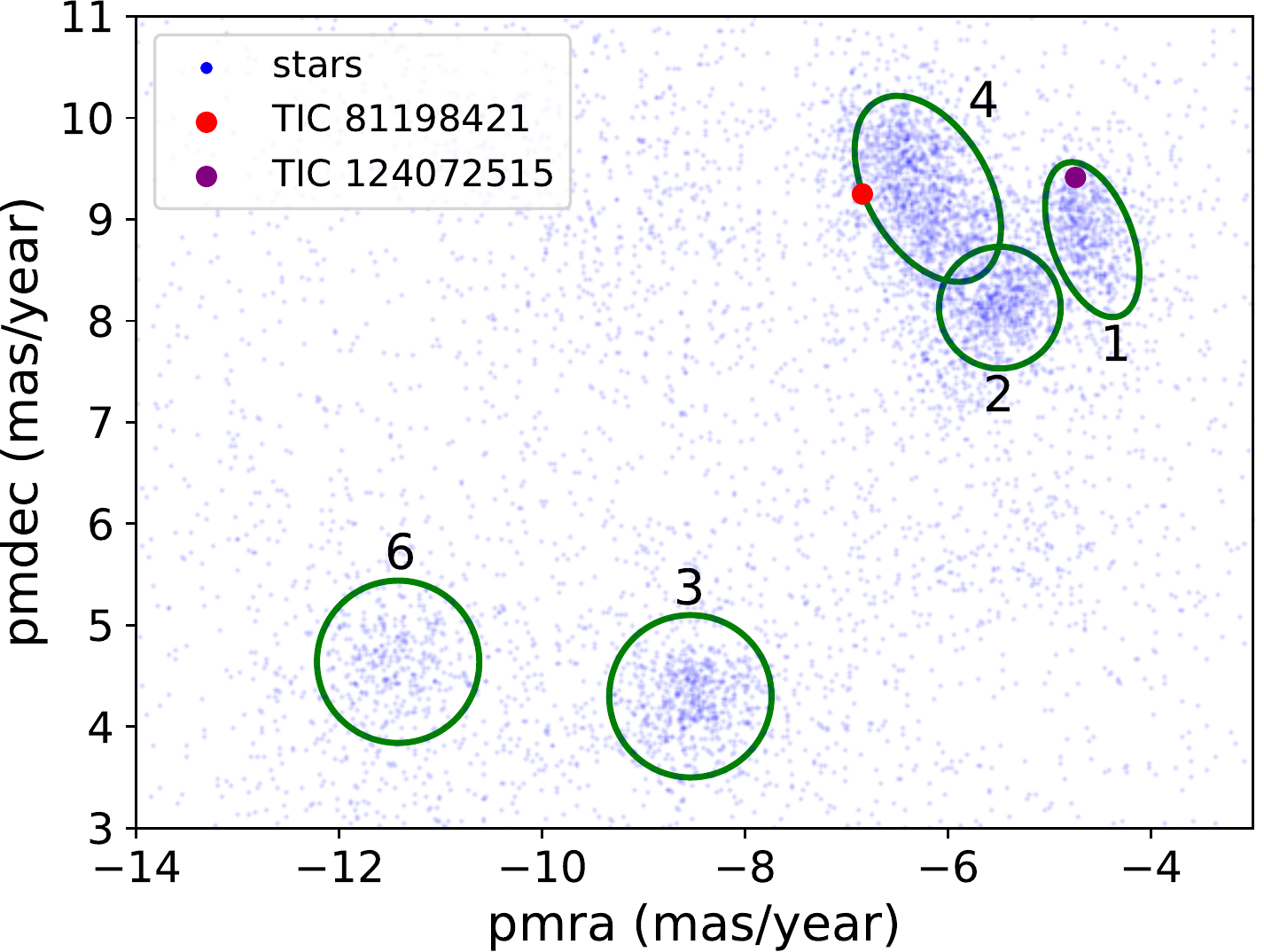}
        \caption{Similar as the left panel of Figure \ref{fig:ScoOB2}, but for Vel OB2. The two dippers are shown by red and purple points. The green ellipses with numbers correspond to approximate cluster regions defined by \citet{2018MNRAS.481L..11B}. Cluster 5 is excluded because of the large variation of the proper motion. \label{fig:VelOB2}}
    \end{center}
\end{figure}

We identified the star clusters proposed by \citet{2018MNRAS.481L..11B} for each dipper in the distribution of the proper motion. Because there are no member catalogs of each cluster, we initially made the distribution of proper motions near Vel OB2 using {\it Gaia} DR2 catalog. The range we used is the right-ascension of $116 < \alpha < 126$ degree and declination of $-50 < \delta < -44$ degree and the range of parallax $\pi > 2$ mas because the average distance of Vel OB2 is much closer than 500 pc. Furthermore, We exclude the sources whose proper motion values were not measured. 

Figure \ref{fig:VelOB2} shows the proper motions of 23,240 stars identified by the procedure described above (blue dots). The green ellipses are approximate areas of the clusters \citet{2018MNRAS.481L..11B} reported. Cluster 5 is excluded because of the large variation of the proper motion. The proper motions of TIC 81198421 and TIC 124072515 are consistent with the memberships of Cluster 1 and Cluster 4. The sky positions and distances are also consistent with these classifications. Using the isochrone analysis, \citet{2018MNRAS.481L..11B} derived the mean ages of Cluster 1 and 4 as about 10 Myr old. This age is also consistent with the presence of the primordial disk in TIC 81198421 and TIC 124072515.


\subsection{Around the Orion}

\begin{figure}
    \begin{center}
        \includegraphics[width=\linewidth]{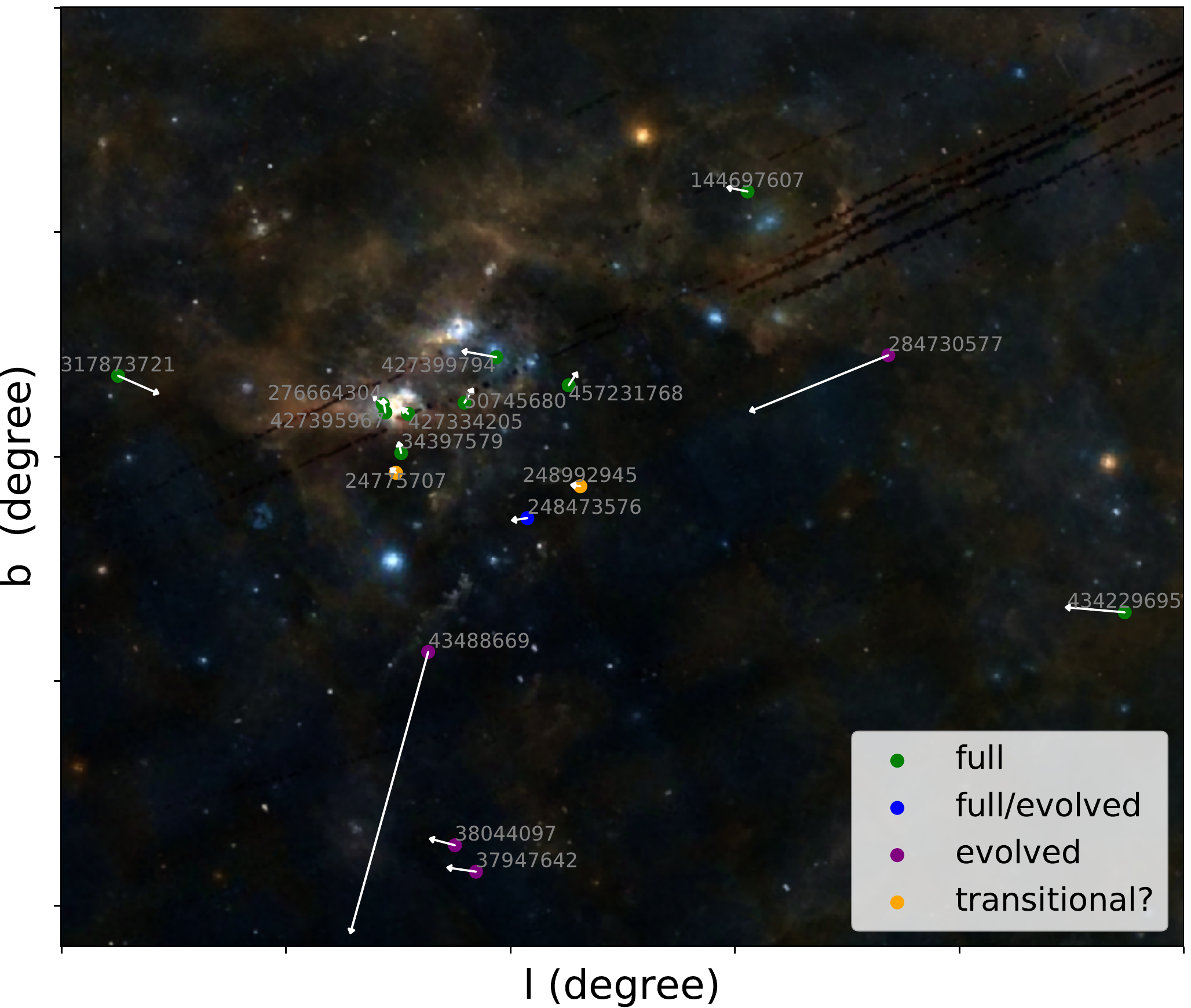}
        \caption{The dippers near the Orion. Each color of dipper represents the disk type classified by the color-color diagram (Fig \ref{fig:disk}). The background picture is synthesized by the images of Digitized Sky Survey  and AKARI.  \label{fig:ori_plot}}
    \end{center}
\end{figure}

We found 17 dippers around the Orion as shown in Figure \ref{fig:ori_plot}. Some of them are close to the Orion molecular cloud complex, while some are relatively far from it.  First, we focus on the following eleven systems, TIC 24775707, 34397579, 50745680, 248992945, 276664304, 457231768, 248473576, 427399794, 427334205, 427395967, and  144697607. These are located near the Orion molecular cloud complex. The remaining six systems, TIC 284730577, 317873721, 37947642, 38044097, 434229695, and 43488669, are relatively far from the Orion molecular cloud complex, which means that these six systems do not belong to the Orion molecular cloud complex.


We note that comprehensive analyses of spectral features will be presented in forthcoming papers while we present several parameters from the spectral analysis for individual objects in this paper.

\subsubsection{Orion complex \label{ss:OC}}

\begin{figure*}
    \begin{center}
        \includegraphics[width=\linewidth]{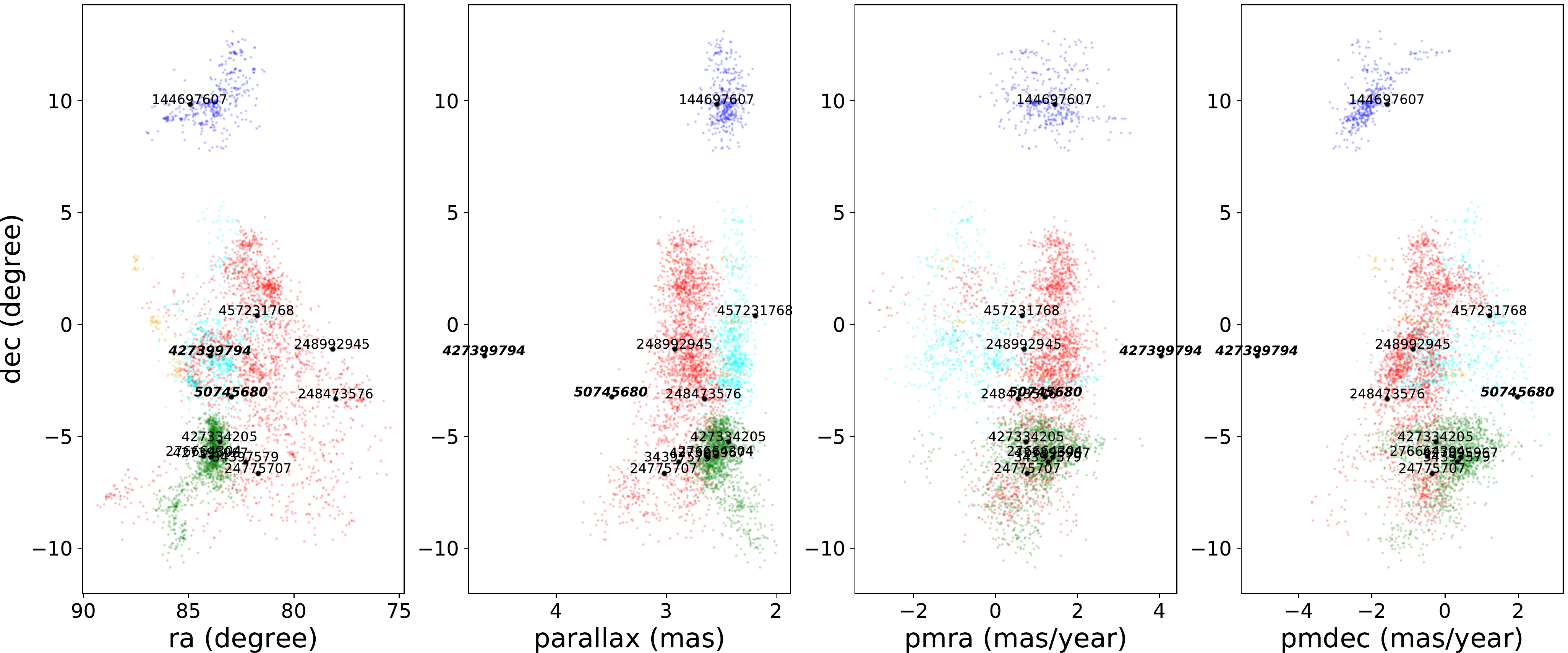}
        \caption{Distributions of positions, distances and proper motions of members of Orion A(green), Orion B(orange), Orion C(cyan), Orion D(red), and $\lambda$ Ori(blue) listed in \citet{2019yCat..51560084K}. TIC 427399794 and 50745680 (bold) do not belong to any clusters. \label{fig:Orion}}
    \end{center}
\end{figure*}

Orion complex is one of the largest star forming regions near the Sun. Kinematic structure of Orion complex was studied by \citet{2018AJ....156...84K} using {\it Gaia} DR2 data. They revealed that Orion Complex had five main kinematic groups, Orion A and B, which are younger parts harboring molecular gases, Orion C and D, which are older parts with little molecular gases, and $\lambda$ Orion, whose gas has mostly been dissipated by a supernova. Based on the population catalog of the clusters \citep{2019yCat..51560084K} (hereafter K19 catalog), we found that TIC 276664304 and 427334205 belongs to Orion A, TIC 457231768 to Orion C, TIC 24775707, 34397579 and 248992945 to Orion D, and TIC 144697607 to $\lambda$ Orion. To identify the memberships of the rest of the dippers, we checked the spatial distributions of the stars listed in the catalogs and the rest of the dippers. Figure \ref{fig:Orion} shows the distribution of the positions, parallaxes and proper motion in the direction of right ascension and declination. The figure shows that TIC 427395967 and 248473576 are likely the members of Orion A and D. However, TIC 427399794 and 50745680 are isolated from the clusters. These two dippers whose membership we could not identify are possibly slowly escaping from their birth place, were born in an isolated small cloud, or have already dissipated birth places. Radial velocity measurements of these dippers will enable further discussion.

We estimated the ages of the nine dippers whose membership we were able to identify. TIC 427395967 and 248473576 are not listed in K19 catalog. We find that the following seven dippers are listed in K19 catalog as the member of sub-groups: TIC 276664304 and 427334205 as ONC-2 ($\sim2.9$ Myr), TIC 457231768 classified as Ori C north-3 ($\sim4.8$ Myr), TIC 34397579 as Ori D south-1 ($\sim3.0$ Myr), TIC 24775707 as Ori D south-5 ($\sim5.0$ Myr), 248992945 as L1616-4 ($\sim7.9$ Myr), and TIC 144697607 as B35-1 ($\sim3.0$ Myr). These results are consistent in that TIC 24775707 and 248992945, classified as relatively older sub-groups, have a transitional disk, whereas the others have a full disk.

\subsubsection{TIC 284730577}

TIC 284730577 (HD 240779) is located at a distance of about 0.5 degree from the dippers in the Orion complex (see Fig \ref{fig:ori_plot}).  This system is a quasi-periodic variable star with infrared excess reported by \citet{2019MNRAS.488.4465G}. This system is also the first field dipper found in the {\it TESS} data. We recover this known dipper by our pipeline. \citet{2019MNRAS.488.4465G} reported that TIC 284730577 was a member of AB Doradus moving group with an age of about 125 Myr, which is consistent with the observation of lithium and rotational period. The equivalent width of the lithium from our observation is consistent with the value (0.12 \AA) reported by \citet{2019MNRAS.488.4465G}.

\subsubsection{TIC 37947642 and 38044097}

TIC 37947642 and 38044097 are located far from any star forming regions near the Orion (Fig.\ref{fig:ori_plot}) except for very wide area of Cassiopeia--Taurus OB association (Cas--Tau; see Fig. \ref{fig:galactic}). Cas--Tau is a young association covering a very large area on the celestial plane and the range of distance is also wide \citep[125--300 pc][]{1999AJ....117..354D}. Cas--Tau is regarded as a physical group and has the same origin as $\alpha$ Persei cluster. Using the membership catalog of Cas--Tau \citep{1999AJ....117..354D} and {\it Gaia} DR2 data, we estimated the tangential velocity distribution of Cas--Tau as $V=25\pm6$ \kms.\footnote{We note that we exclude HIP 2377 from Cas-Tau member because its parallax is much smaller than other members.} Although the sky positions of TIC 37947642 and 38044097 are in the Cas--Tau region, these are likely not to belong to Cas--Tau because of much smaller tangential velocities ($\sim5$ \kms) compared with those of Cas--Tau. 

The distances of TIC 37947642 and 38044097 are much larger than 100 pc. Because all the well-known young moving groups are located at less than 100 pc, we cannot identify the memberships of them. However, the dimming events with the IR excess supports that TIC 37947642 and 38044097 are young stars with circumstellar disks. The fact that their proper motions go toward almost the same direction and their sky positions are very close supports that TIC 37947642 and 38044097 could belong to a unknown common young moving group. 

\subsubsection{TIC 43488669}
TIC 43488669 is far from any star forming regions except for Cas--Tau as shown in Figure \ref{fig:ori_plot}. Because its tangential velocity is very high ($\sim55$ \kms), this system surely does not have Cas--Tau membership. While we detected an absorption feature of a H$\alpha$ line, we also detected the lithium absorption line. The spectrum also exhibits the high radial velocity of $\sim46$\kms. Its high velocity supports the fact that TIC 43488669 is a runaway star. We will revisit this system in Section \ref{sec:runaway}.

\subsubsection{TIC 434229695 and 317873721}
TIC 434229695 is also far from Orion molecular cloud complex and Taurus molecular cloud as shown in Figure \ref{fig:ori_plot}. {\it Gaia} measured the radial velocity of TIC 43422969, so we can calculate its galactic UVW space motion. The galactic UVW space motion, $(UVW) = (-8.93, -11.51, -3.64)$\kms does not matche to any young moving groups and associations even to Cas--Tau ($(UVW)=(-13.24, -19.69, -6.38)$\kms). In addition to the IR excess, the high-resolution spectrum shows a clear H$\alpha$ emission and also a lithium absorption, supporting the fact that TIC 43422969 is a young stellar object with an accreting disk. TIC 43422969 has also been identified as the T-Tauri star candidate based on the infrared satellite survey, AKARI \citep{2010A&A...519A..83T}.

TIC 317873721 is far from any star forming regions. We detected an inverse P-Cyg type of H$\alpha$ emission line. However the temperature is so high ($\sim 8,000$ K as shown in Table.\ref{tab:stellar_properties}) that we do not detect the lithium absorption line, which is one of the clues to determine its age. We will present the detailed analysis of the spectral features in a forthcoming paper.


\subsection{Dippers in the Field}
Three dippers (TIC 57830249, 47390297 and 167303776) are located far from any nearby molecular clouds and OB associations. We tried to identify their memberships in the following sections.

\subsubsection{TIC 57830249}
TIC 57830249 was previously reported as TWA 33 by \citet{2012ApJ...757..163S}. They reported that TWA 33 was a member of TW Hydrae Association (TWA) based on its position, proper motion, spectral signature, distance estimate, and a mid-IR excess. TWA is the nearest association of very young low-mass stars (usually classified as T association), whose distance is within 100pc. TWA 33 exhibits the infrared excess \citep{2012ApJ...757..163S}, but optical variability has not been reported yet. The age of TWA 33 was estimated as $10\pm3$ Myr \citep{2015MNRAS.454..593B}, being consistent with a primordial disk structure. \citet{2012ApJ...757..163S} measured equivalent width of H$\alpha$ emission line (EWH$\alpha$ = $5.8$\AA) and spectral type (M$4.7\pm0.5$) with the spectroscopic observation. According to \citet{2003ApJ...582.1109W}, who proposed empirical criteria for distinguishing classical T-Tauri stars (CTTSs) and weak-line T-Tauri stars (WTTSs) using equivalent width of H$\alpha$ emmision line, TWA 33 is classified as WTTS. Our classification of TIC 57830249 as evolved system supports this results. This indicates that TIC 57830249 has no accretion signature.

\subsubsection{TIC 47390297 \label{ss:anotherdip}}
TIC 47390297 is quite far from any star forming regions. Its large distance (676 pc) prevents us from identifying membership of any young moving groups. From {\it Gaia} DR2 data, TIC 47390297 has very high proper motion ($V=\sim40$\kms). This velocity is above galactic deviation near the Sun ($\sigma V_{gal}=\sim30$\kms), TIC 47390297 is possibly a runaway star.

\subsubsection{TIC 167303776}

TIC 167303776 is located near the south ecliptic pole and also far from any star forming regions. The direction of proper motion is different from the other dippers close to Sco OB2, which implies TIC 167303776 is not a member of Sco OB2. Using its position, parallax and proper motion data provided {\it Gaia} DR2, we examined possibility of any young moving group by BANYAN $\Sigma$ \citep{2018ApJ...862..138G}. BANYAN $\Sigma$ is an online Bayesian classifier to statically determine the memberships of stars to a nearby young moving group. Based on the stellar position, parallax and proper motion, BANYAN $\Sigma$ computes the membership probability of each young moving group even without radial velocity data.

From the results of BANYAN $\Sigma$, TIC 167303776 has a membership of Carina association with $80.4\%$ probability, as well as field probability is $19.3\%$ and a probability of Carina-near association moving group is $0.4\%$. Because its probability are less than $90\%$, \citet{2018ApJ...862..138G} have not reported TIC 167303776 as a Carina member. Because radial velocity data is not yet available, we do not fully identify its membership although the infrared excess indicates TIC 167303776 is young, and it is consistent that TIC 167303776 belongs to the young moving group. However, TIC 167303776 is potentially a member of Carina association with a mildly high probability.

\citet{2015MNRAS.454..593B} calculated the isochronal ages for young moving groups in the solar neighbourhood. They determined the age of Carina association as $45^{+11}_{-7}$ Myr. According to the discussion so far, the age of TIC 167303776 is at least 38 Myr. Comparing the classical dissipation time scale of primordial disk ($\sim10$Myr), TIC 167303776 is much older and it is unlikely to still have a primordial disk structure. Like HD 240779, the infrared excess and dimming event of TIC 167303776 is potentially explained by collisions of planetesimals in the debris disk.


\section{Runaway old dipper: TIC 43488669} \label{sec:runaway}

Among 35 dippers that we found, TIC 43488669 is of particular interest; it was subjected to follow-up observations because of its large proper motion and radial velocity. Moreover, the infrared excess indicates that TIC 43488669 is between class I\hspace{-.1em}I and class I\hspace{-.1em}I\hspace{-.1em}I. The age of TIC 43488669 is much older than that of most dippers found near the star forming regions. In this section, we discuss the runaway old dipper, TIC 43488669.

\subsection{Light curve of TIC 43488669}

\begin{figure}
 \begin{center}
     \includegraphics[width=\linewidth]{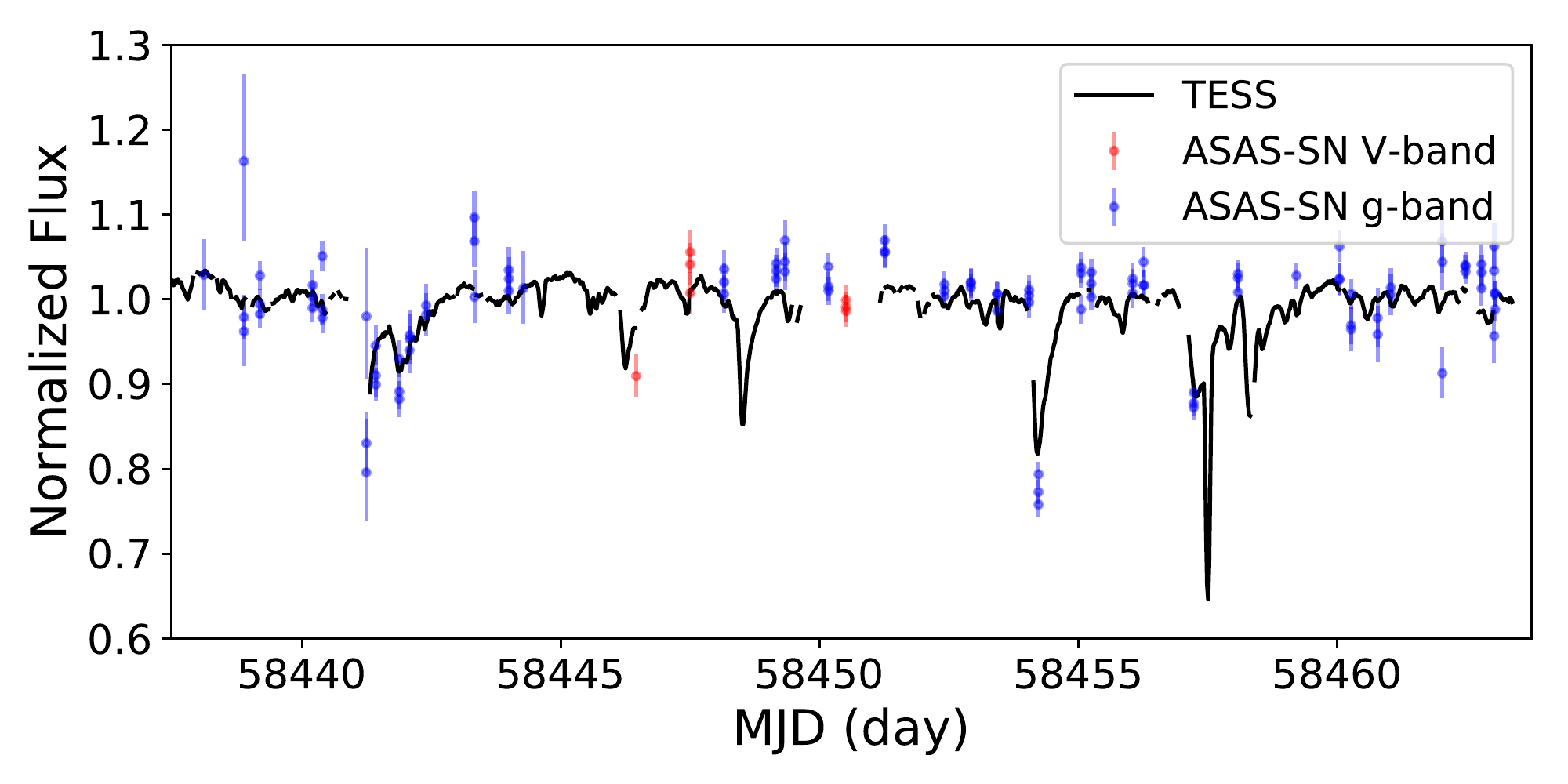}
     \caption{Light curves of TIC 43488669 from {\it TESS} (solid line) and ASAS-SN (error bars). \label{fig:43488669lc}}
 \end{center}
 \end{figure}

The light curve of TIC 43488669 shown in Figure \ref{fig:43488669lc} exhibits several asymmetric dips with tails. The tail feature is similar as ones found in the evaporating planet candidates KIC 12557548b \citep[e.g.][]{2012ApJ...752....1R,2013ApJ...776L...6K} and exocomets \citep[e.g.][]{2018MNRAS.474.1453R,2019A&A...625L..13Z} but the depth of dips TIC 43488669 shows is much deeper an order of magnitude than both. The tail in  TIC 43488669  is also likely due to a dusty tail by radiation pressure. However, further study is required for unraveling what produces the tail. We also confirm that the dips are in the ASAS-SN light curve \citep{2014ApJ...788...48S,2017PASP..129j4502K}. The ASAS-SN light curves well agree with the {\it TESS} light curve and we found that TIC 43488669 continuously exhibits such dips at least in several year scale. As will be explained in the following subsections, from the lithium line and IR excess, this system is likely to have a debris disk and the age is about 100--400 Myr. Considering the tail feature and no gas disk, the dips are likely to be produced by asteroids or debris in a disk.





\subsection{Stellar Properties}

\begin{figure*}
    \begin{center}
        \includegraphics[width=\linewidth]{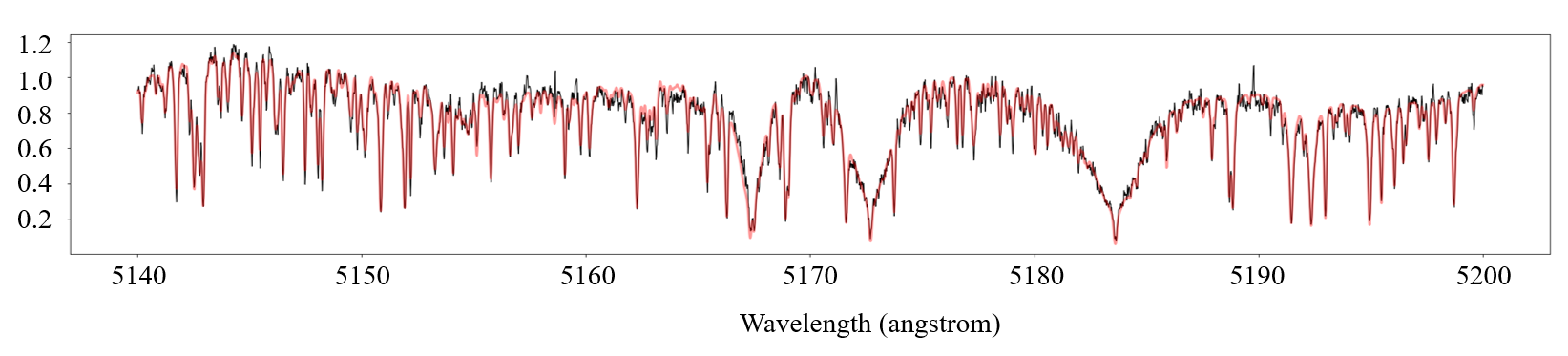}
        \caption{The high-dispersion spectrum of TIC 43488669 around the Mg-triplet (black). The red curve is the best-fit model by specmatch-emp \citep{2017ApJ...836...77Y}. \label{fig:hds}}
    \end{center}
\end{figure*}

We obtained stellar parameters by fitting the stellar model to the high dispersion spectrum around the Mg triplet by Subaru/HDS \citep{2002PASJ...54..855N} as shown in Figure \ref{fig:hds}. The parameters were derived using the public code, specmatch-emp \citep{2017ApJ...836...77Y}, the stellar temperature $T_\mathrm{eff} = 5080 \pm 100$ K, The stellar radius = 0.71, and the metallicity of [Fe/H]= $-0.53 \pm 0.09$. Also, the radial velocity is $46 \pm 1 $ \kms. We also measure the equivalent width (EW) of the lithium 6,708.6 \AA\, line with the Gaussian model and obtain $\mathrm{EW_{Li}} = 56\pm4$ m\AA\,  from the spectrum of TIC 43488669. Following \citet{2019MNRAS.488.4465G}, we estimated the age of the system using the EW of lithium based on known young clusters near the Sun. Measurement of Hyades, UMa and Pleades are from \citet{1993ApJ...415..150T}, and \citet{2009A&A...508..677A} \citet{2016A&A...596A.113B}. Measurement of other clusters are from \citet{2008ApJ...689.1127M}. 
Compared with the stars associated with known clusters, the age of TIC 43488669 is estimated as 100--400 Myr. We note that the depletion rate of lithium in pre-main sequence stars is strongly dependent on metallicity
while the depletion rate in the main sequence star is insensitive to composition: Metal-poor pre-main sequence stars deplete more slowly than metal-rich ones \citep{2014ApJ...790...72S}. In this sense, TIC 43488669 is potentially older than 100--400 Myr because of its poor-metalicity. However, it is difficult to estimate the age with consideration of this effect. Therefore, we assume 100--400 Myr as a minimal value in the following discussion. 


\begin{figure}
    \begin{center}
        \includegraphics[width=\linewidth]{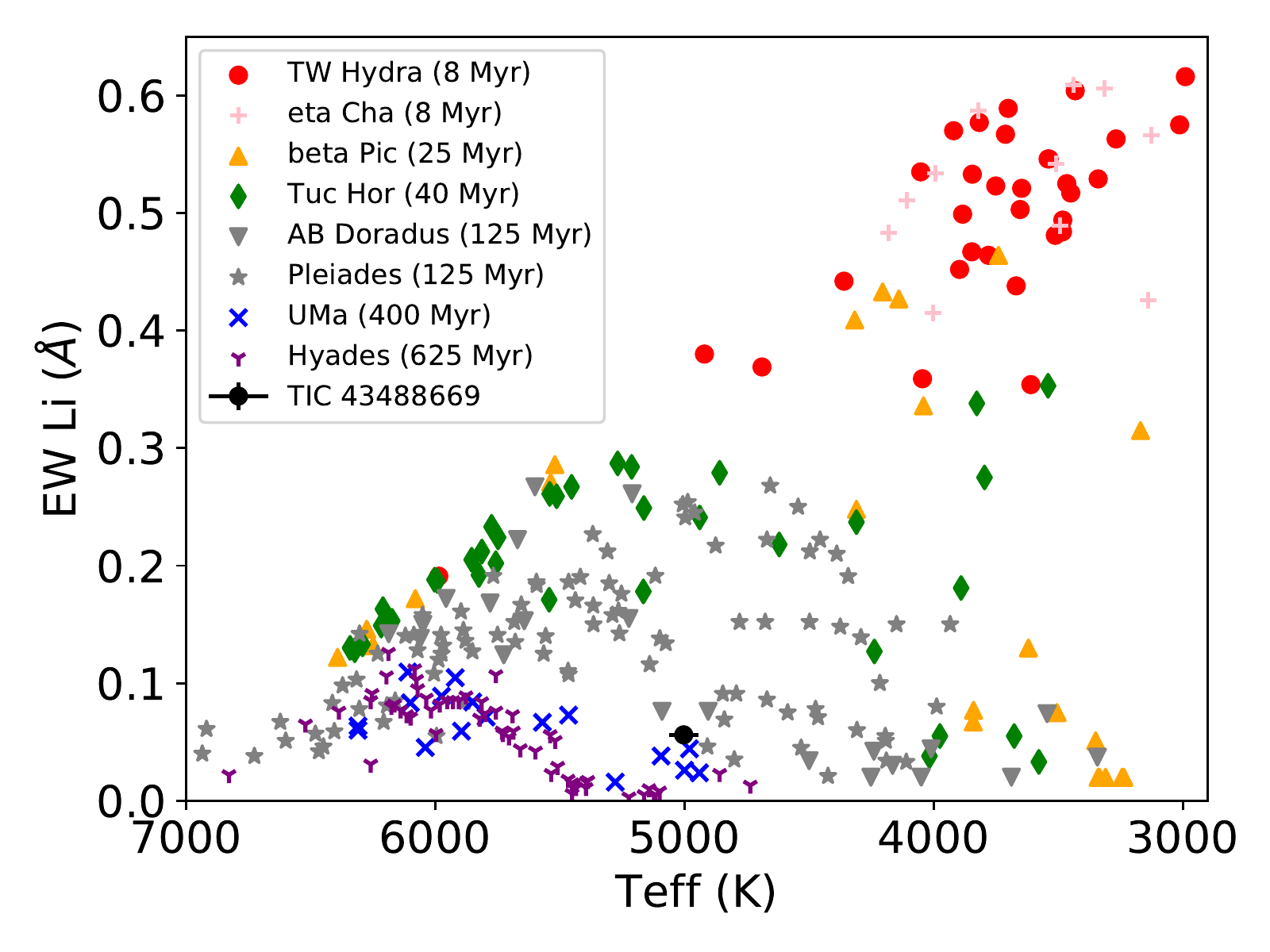}
        \caption{The equivalent width of lithium 6707 \AA\, line, following \citet{2019MNRAS.488.4465G}. The black dot with the error bar indicates TIC 43488669. \label{fig:Teff-EWLi}}
    \end{center}
\end{figure}

\subsection{Orbit from 3 dimensional velocity}

We noted that {TIC 43488669} shows a large 3D velocity ($\sim 78 \kms$) relative to the Sun. 
To understand the origin of this star, 
we computed its orbit in the past. First, we evaluated the heliocentric position and velocity of {TIC 43488669} by using the astrometric data from {\it Gaia} and the measured radial velocity of $v_\mathrm{los}=56 \kms$. 
Given the high-quality data, we did not take the observational error into account. 
Then, we estimated the Galactocentric positon and velocity of this star 
to be $(x,y,z)=(-8.32, -0.11, -0.12) \kpc$ 
and $(v_x, v_y, v_z)=(-14.87, 214.96, -74.56) \kms$, respectively,
by assuming that the Galactocentric position and velocity of the Sun is given by 
$(x_\odot, y_\odot, z_\odot) = (-8.112, 0, 0) \kpc$ \citep{Gravity2018} and 
$(v_{x,\odot}, v_{y,\odot}, v_{z,\odot}) = (11.1, 245.31, 7.25) \kms$  \citep{2004ApJ...616..872R, 2010MNRAS.403.1829S}, respectively. 
Based on the current position and velocity of this star, we computed the orbit of this star in the last $300$ Myr in a realistic model potential of the Milky Way given by \cite{2017MNRAS.465...76M}. 

We note that, if we assume that the Local Standard of Rest (LSR) velocity is $233.09 \kms$ 
(which corresponds to the circular velocity of our model potential at the Sun), 
the 3D velocity of {TIC 43488669} relative to the LSR is 
$(U, V, W) = (-14.87, -18.13, -74.56) \kms$.

\begin{figure*}
    \begin{center}
        \includegraphics[width=0.45\linewidth]{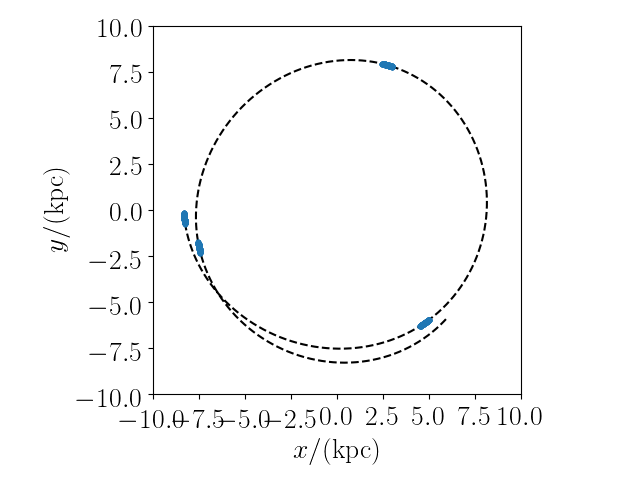}
        \includegraphics[width=0.45\linewidth]{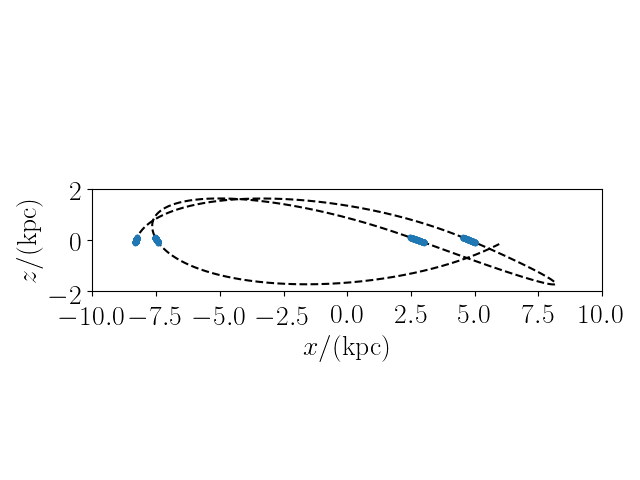}
        \caption{Orbit of TIC 43488669 in the last 300 Myr. The left and right panels indicates the trajectory on the Galactic $(x,y)$-plane and $(x,z)$-plane. The current location of the  Sun corresponds to $(x_\odot, y_\odot, z_\odot) = (-8.122, 0, 0) \kpc$. 
        To guide the eye, the disk-crossing locations are marked with blue thick region. \label{fig:velo}}
    \end{center}
\end{figure*}

The orbit in the Galactocentric $(x, y)$- and $(x,z)$-planes is shown in Figure \ref{fig:velo}. 
We note that this star shows a nearly circular orbit in $(x,y)$-plane with an eccentricity of $e=0.06$, but shows a large excursion above and below the Galactic plane with the maximum excursion of $z_\mathrm{max} = 2 \kpc$. 
This large excursion is unusual given the young age of this star ($\sim 100$--$400$ Myr from the Lithium line), 
which hints that this star is a runaway star ejected from its birthplace. 
The orbit passes through the Galactic plane every $\sim 75$ Myr. 
Because of the near-circular nature of the in-plane motion, the vertical motion of this star is well decoupled from its in-plane motion. 
Consequently, this star passes through the disk plane with a velocity of $\sim 80 \kms$ relative to the local circular velocity in the last $300$ Myr. 
This indicates that, if this star is a runway star ejected from the Galactic plane, its ejection velocity can be estimated to be $\sim 80 \kms$. 
This estimate of the ejection velocity is not sensitive to the age of this star or when it was ejected as long as the ejection happened at the disk plane. 

If we assume that the system was ejected from a star cluster near the Galactic plane (indicated by blue), the ejection velocity should be $\sim 80 \kms$ toward almost $z$-direction. Such a high velocity is expected by dynamical ejection of stars from star clusters \citep{2012ApJ...746...15B}. Thus, the large ejection velocity supports the scenario that TIC 43488669 is a runaway star ejected from its birth cluster a few hundred Myr ago. 
 
\section{Discussion}

\subsection{The dippers whose memberships are unidentified}

 \begin{table}
 \caption{The dippers whose memberships are not identified in the paper. \label{tab:unidentify}}
 \begin{center}
 \begin{tabular}{ccccc} 
  \hline
  \hline
 TIC ID & PM$>30$\kms & Disk & H$\alpha$ & Origins\\
\hline
43488669 & \checkmark & E & absorption & runaway\\
47390297  & \checkmark & F & N/A & runaway?\\
38044097 & & E & N/A  & moving group? \\
37947642 & & F/E & N/A  & moving group? \\
50745680 & & F & N/A  & close to SFR\\
427399794 & & F & N/A & close to SFR\\
317873721 & & F & emission & \\
434229695 & & F & emission & \\
 \end{tabular}
 \end{center}
 \end{table}

In the previous section, we showed a possible scenario of the origin of the runaway dipper, TIC 43488669: The system was ejected from a birthplace, likely located at a star cluster on the galactic plane in several Myr ago. From the IR excess and the age estimated from the lithium line, we determined that it was the debris disk part of its evolution stage. Considering the evolution stage and the tail-like features in the light curve, the dips were likely produced by dust, asteroids or material formed in the debris disk. 

Besides TIC 43488669, the membership for seven dippers was not associated with known nearby star forming regions, and moving groups in this paper. Table \ref{tab:unidentify} summarizes their properties. Some of them (TIC 38044097 and 37947642) will potentially remain unidentifiable due to lack of the catalogs of moving groups. However, the rests are likely stars that exists outside of any star forming regions, or moving groups. 

The ejection mechanism discussed in \S 4 can explain the escape of such systems from their birthplace. TIC 47390297, which exhibits high proper motion ($\sim 40$ \kms), might be a runaway dipper. Also, TIC 50745680 and 427399794 are located near the Orion clusters and have small proper motions. The sky positions and distances of these two dippers can be attributed to the slow escape from their birthplaces. However, TIC 434229695 and 317873721, which also have small proper motion but clearly show H$\alpha$ emission lines, cannot be attributed to a runaway process from their birthplace because their location are far from any nearby star forming regions. A birth in a small clump like ``Bok globule'' or `` quick'' dissipation of clouds can explain their origins. These results are also indicative of the YSO distribution because only parts of the T-Tauri stars exhibit dips in their light curve. Our results imply that it is likely that there are numerous YSOs outside the star forming regions, which have been overlooked thus far. The IR excess with astrometric information by {\it Gaia} DR2 can be used to find such isolated YSOs. Indeed, TIC 43422969 was already been identified as a T-Tauri star candidate, from the analysis of the infrared catalog by AKARI \citep{2010A&A...519A..83T}, located outside the star forming regions.   

\subsection{Occurrence of the debris dipper in field}

 In our survey, we discovered three old dippers (TIC 43488669, 167303776 and 284730577) whose age are $>10$ Myr. Since the typical dissipation timescale of disk is 10 Myr, these systems are unlikely to have a primordial disk and the IR excess originates from a secondary disk, namely a debris disk. We call such stars ``debris dipper'' here although the term of ``dipper'' is usually used for a pre-main sequence star whose light curve exhibits dips.


\begin{figure*}
    \begin{center}
        \includegraphics[width=\linewidth]{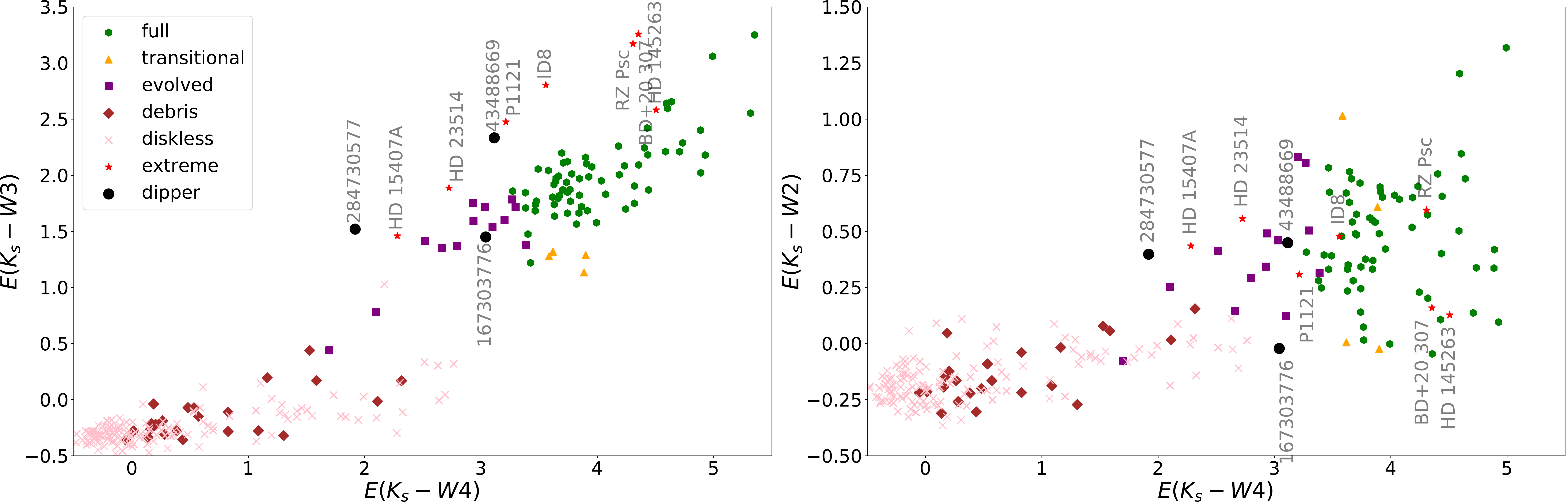}
        \caption{IR color-color diagrams for classification of the disk type similar as the Figure \ref{fig:disk}. The newly plotted red points are the extreme debris disks reported by \citet{2015ApJ...805...77M} and RZ Psc. The left panel shows the IR color at the $W3$ band relative to the $W4$ band and the right panel indicates the IR color at the $W2$ band relative to the $W4$ band. All  of the systems exhbit a strong IR excess relative to the other debris disks (brown diamonds). \label{fig:disk_extdeb}}
    \end{center}
\end{figure*}

Figure \ref{fig:disk_extdeb} shows the debris dippers on the color-color diagram, same as Figure \ref{fig:disk}. These debreis dippers have a similar excess as that observed in the evolved disks. However, the debris dippers cannot be classified as the evolved or full disk systems owing to their old ages. It is known that about $1\%$ of the debris disk has a high infrared excess due to a warm circumstellar dust \citep{2009ApJ...698.1989B, 2013prpl.conf2B068K}. Such debris disks are dubbed ``extreme debris disks'', defined by the star whose disk luminosity is roughly 1 \% larger than the stellar luminosity. Then, the IR excess of these three debris dippers should be classified as the extreme debris disk (see section \ref{sec:result}). We also plot six known extreme debris disk hosts, P1121 (80 Myr), HD 15407A (80 Myr), HD 23514 (120 Myr), HD 145263 (10 Myr), BD+20 307($\gtrsim 1$ Gyr), and ID8 (35 Myr) listed in \citet{2015ApJ...805...77M}, and RZ Psc which is one of the extreme debris disk candidates \citep{2013A&A...553L...1D}. We also compute a $W2$ color excess ($E(K_s-W2)$) by the same procedure described in Section \ref{sec:result}, and plot them against $W4$ color excess. All the extreme debris disks and the debris dippers exhibit a strong IR excess at $W3$ and $W4$ bands, compared to the regular debris disk hosts (brown diamonds) regardless of their ages.  At the $W2$ band, most of the extreme debris disks and debris dippers have large dispersion, but, exhibits a larger excess than those of the regular debris disks. These results suggest that the debris dippers we found should be classified as the extreme debris disk. TIC 167303776 exhibits a relatively weak IR excess at $W3$  respect to $W4$ compared to the other extreme debris disks, which is possibly owing to the weak dust heating; the effective temperature of TIC 167303776 listed in TICv8 (3149 K) is lower than any other systems ($>5000$ K). 

\citet{2019AJ....157..202S} reported the variability in infrared starlight of the two extreme debris disks, ID8 and P1121. However photometric variability of the extreme debris disk in the optical band is rare. RZ Psc was believed to be the one of extreme case but its revised age ($20_{-5}^{+3}$ Myr; \citet{2019A&A...630A..64P}) challenges the claim that the system has a debris disk. The revised age rather supports that the IR excess originates from a protoplanetary disk. Another example is HD 240779 (TIC 284730577) \citep{2019MNRAS.488.4465G}, which is also identified as the dipper by our pipeline. The high infrared excess with the age of $125$Myr old supports that HD 240779 has an extreme debris disk. They suggested that an occultation of a dust cloud as an origin of dips, which is probably produced by a rapid disintegration or evaporation of planetesimals. However the scenario of dimming event is poorly understood. To constrain the dimming scenario, the occurrence of such a debris dipper is required.

Here, we derive the occurrence of the debris dipper to test the consistency with the geometric occultation probability of the dust clouds. First, we estimated the number of the systems in CTLv8 whose age is less than 500 Myr. CTLv8 consists of a combination of main-sequence stars whose magnitude in the {\it TESS} band is lower than 13 and the stars selected from the Cool Dwarf List \citep{2019AJ....158..138S}. Therefore, CTLv8 exhibits a bimodal mass distribution. We divided the stars in CTLv8 into high- and low-mass groups by the threshold of $M = 0.68M_{\odot}$. The median masses of the two groups are 1.18$\mathrm{}{M_{\odot}}$ and 0.53 $\mathrm{}{M_{\odot}}$. Then, we estimated a mean lifetime $\tau_{ms}$, 5.2 Gyr for the high-mass group and 121 Gyr for the low-mass group, using the following relationship
\begin{equation}
\tau_{ms} = \frac{M}{L} \approx 10^{10}\left(\frac{M}{M_{\odot}}\right)^4.
\end{equation}
For the high-mass group, we estimated the number of the stars whose age is less than 500 Myr assuming a uniform distribution from 0 to 5.2 Gyr. However, the mean lifetime of low-mass group is much larger than the age of the universe. Instead, we assumed a uniform distribution of  the stellar age from 0 to 10 Gyr based on the age distribution near the sun \citep{2019arXiv190607489R} to derive the number of the stars with $<500$ Myr. From this estimate, the approximate number of the stars whose age is less than 500 Myr is $4\times10^{5}$. 

The occurrence ratio of extreme debris disks to young post-main sequence stars ($<120$ Myr) is estimated as about $1\%$ \citep{2013prpl.conf2B068K}. Assuming that this ratio is constant up to 500 Myr, we derived the approximate number of the extreme debris disk in CTLv8 as $4\times10^{3}$. 

We assumed that we did not overlook most of the dimming events by field stars in our pipeline. Note that we are likely to overlook some dippers in star forming regions because stars in dense clusters are highly contaminated by nearby stars, which obscures their dimming events. The contamination by nearby stars is less severe for the stars in the filed than those in star forming regions. The three debris dippers in $4\times10^{3}$ extreme debris disks means that the probability of variability in extreme debris disks is about $p \sim 0.1\%$.

In addition, we assumed that all the occultation in extreme debris disks are caused by dust clouds and each dip is caused by a single dust cloud in a circular orbit. The semi-major axis of a dust cloud with a half of transit duration $d_\mathrm{half}$ is written as 
\begin{equation}
\label{eq:aGM}
a = \frac{G M d_\mathrm{half}^2}{R^2},
\end{equation}
where $G$ is the gravitational constant, $M$ and $R$ is the stellar mass and radius. Using equation (\ref{eq:aGM}), the transit geometric probability $p_\mathrm{geo} \approx R/a$ is estimated as 
\begin{equation}\label{eq:p_geo}
p_\mathrm{geo} \sim 0.01 \left(\frac{\rho}{\rho_\odot}\right)^{-1} \left(\frac{d_\mathrm{half}}{4 \, \mathrm{hour}}\right)^{-2}, 
\end{equation}
where $\rho$ is the stellar density and $\rho_\odot$ is the solar value of $\rho$ and we adopt the typical half duration time of the dip of $4\pm1$ hour from the light curve. Denoting the occurrence of clouds which are much dense and clustered for obscuring the stellar photosphere in extreme debris disk by $p_c$, we obtain $p_c = p/p_\mathrm{geo} \sim 0.1$. At least, because $p_c$ does not exceed unity, we cannot reject the cloud scenario as the origin of the debris dipper. We note that several dippers locating in Sco OB2, whose age is estimated as $\sim 11-17$ Myr \citep{2012ApJ...746..154P}, unlikely host a protoplanetary disk, but an (extreme) debris disk, which might increase $p_c$.
Note that the equation.\ref{eq:p_geo} is assumed that a dust clump is much smaller than the radius of the star. Assuming a clump with a scale similar to or larger than the star radius, $p_c$ can be larger than $0.1$.

\subsection{Possible scenario of the debris dipper}
All of the debris dippers we found exhibit a strong IR excess at the $W3$ and $W4$ bands and the probability of variability in the extreme debris disks is about $p \sim 0.1\%$, which is consistent the geometrical probability of the transiting dust cloud. This fact indicates that it is feasible that the dimming event of the debris dipper is caused by the occultation of the dust clouds or similar structures locating at the inner disk. The question is the origin of the dust clouds orbiting around the star. 

The key feature to unravel the origin of dust clouds is periodicity. \citet{2019MNRAS.488.4465G} revealed that HD 240779 has two peaks in the Lomb-Scargle periodogram of the light curve: one is located at about 4.2 day of the rotational period and another one is found at 1.51 day of the dimming period. They focused on the dimming periodicity and suggested that the occultation was caused by a transiting clustered dust cloud orbiting at a distance of 6 stellar radii. We examined if their scenario could be applied to the other two debris dippers.

\begin{figure}
    \begin{center}
        \includegraphics[width=\linewidth]{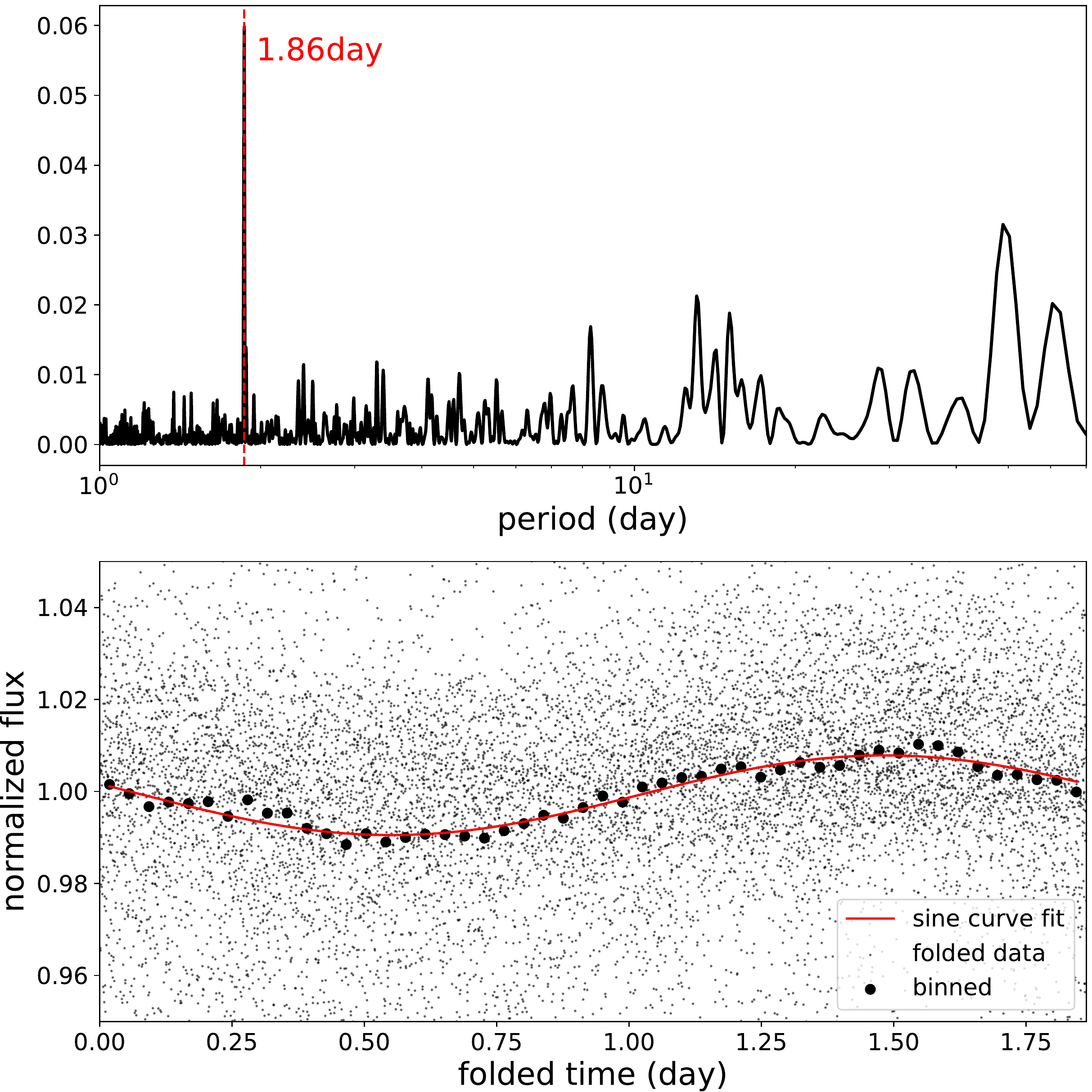}
        \caption{Upper --- The Lomb Scargle periodogram of the entire sector of TIC 167303776 except for sector 7 (black solid line). This periodogram exhibits a single clear peak at 1.86 day (indicate by the red dashed line). Bottom --- The small dots show the light curve of TIC 167303776 folded by 1.86 day. The large dots show the binned flux. The red curve is the best-fit sine function of the binned flux. The folded flux is well expressed as the sine curve, indicating that the 1.86 day signal is likely to be owing to the rotational period. \label{fig:167303776_period}}
    \end{center}
\end{figure}

TIC 167303776, which is located near the south ecliptic pole, have been monitored for the entire sectors of {\it TESS} one year's mission except for sector 7. The Lomb-Scargle periodogram \citep{1982ApJ...263..835S} of the entire light curve exhibits the only single clear peak at 1.86 day (as shown in the upper panel of Figure \ref{fig:167303776_period}). Because the light curve folded by 1.86 d is well approximated by a sine curve (the lower panel of Figure \ref{fig:167303776_period}), this signal is likely to originates from the stellar rotation. The absence of the other peaks in the periodogram indicates that the dimming event is not (quasi) periodic but aperiodic in contrast to HD 240779. 

\begin{figure}
    \begin{center}
        \includegraphics[width=\linewidth]{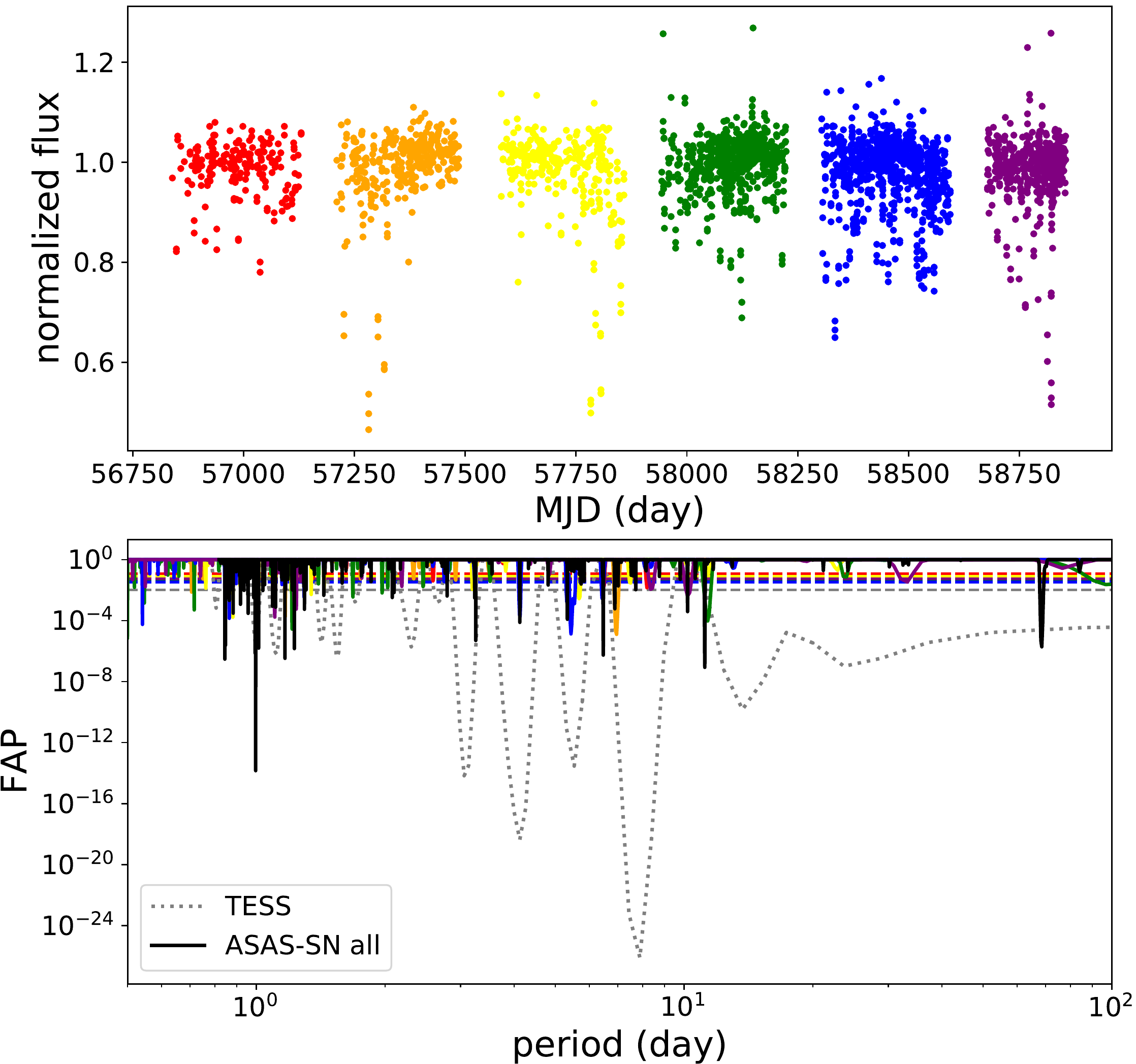}
        \caption{Upper --- The ASAS-SN light curves of TIC 43488669 in six years. Both $V$ and $B$ magnituded are converted into the normalized flux and are plotted together in the same panel, ignoring the wavelength dependence of the dimming. We devided the ASAS-SN light curve into six terms with the different colors. Bottom --- The Lomb Scargle periodogram of the TIC 43488669. The solid colored lines show the periodograms in terms corresponding the color of the upper panel. The solid black line corresponds to the entire light curve of ASAS-SN and the gray dotted line corresponds to the {\it TESS} light curve. The dashed line shows the false alert probability levels corresponding to each color.} \label{fig:43488669_period}
    \end{center}
\end{figure}


TIC 43488669 was monitored only in one sector of the {\it TESS} first one year observation. Instead, ASAS-SN has monitored this system in about six years and provides plenty of photometric data although 167303776 is a bit faint ($\mathrm{Tmag} = 14.5$) for the ASAS-SN data. The upper panel of Figure \ref{fig:43488669_period} shows the ASAS-SN light curves in the six years. We converted both $V$ and $B$ magnitudes into the normalized flux and plotted together, ignoring the wavelength dependence of the dimming. The bottom panel shows the Lomb Scargle periodogram.  Although the periodogram of the entire ASAS-SN light curve exhibits several peaks for each term, no common peak is found in the entire light curve. The periodogram indicates that the dimming period changes during the ASAS-SN observation or the dimming events are episodic. 

\begin{figure*}
    \begin{center}
        \includegraphics[width=\linewidth]{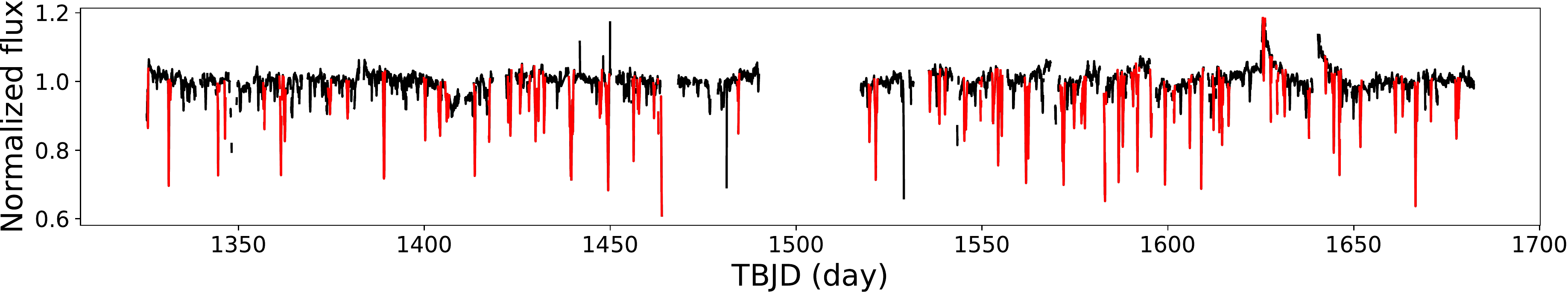}
        \caption{The light curves of TIC 167303776 monitored in the entire sector except for sector 7 (red and black lines). The red lines indicate the dips whose depth are larger than 10 \% against the flat level. Seventy three large dips are contained in the entire light curve. \label{fig:167303776_dip}}
    \end{center}
\end{figure*}

We did not find clear periodicity both in TIC 43488669 and 167303776. An alternative scenario to the one \citep{2019MNRAS.488.4465G} proposed is required to account for their dips. The simple explanation is that those are transiting events by dust clumps randomly distributed in the orbits. To make a dust clump, one needs to assume a planetesimal or embryo which traps dust in their Hill radius. From here, we discuss the case of TIC 167303776 whose radius and mass is, according to TICv8, 0.35 times of the Sun. A dust clump with a radius of 8,000 km is required to obscure the area more than 10\% of the stellar disk. Assuming that a planetesimal or embryo core envelopes dust all over its Hill radius, a planetesimal whose mass is larger than $3 \times 10^{20} \mathrm{kg}$ can form such a dust clump at $1 \mathrm{AU}$, which is the preferred distance derived by mean transit duration of TIC 167303776 (see Equation \ref{eq:aGM}). We note that the mass of the planetesimal core required to be the 8,000km dust clump is inversely proportional to the cube of the semi-major axis. The next question is how many dust clumps are required to account for the dimming events, which are observed in the entire sector of {\it TESS} observation. One the author (T.T.) visually inspected the number of the dips in the entire light curve of TIC 167303776 and found 72 dips whose depth are larger than 10\% against the flat level (Figure \ref{fig:167303776_dip}). A period of a dust clump located at 1 AU is about 1.7 year in the case of TIC 167303776. Therefore, to explain the entire light curve by transiting dust clumps, at least a hundred of $\gtrsim 3 \times 10^{20} \mathrm{kg}$ planetesimals are required, which corresponds to a $300 \mathrm{km}$ radius assuming a bulk density of $3 \mathrm{g cm^{-3}}$. The total mass of planetesimals is roughly consistent with the minimum mass solar nebula about 1--2 AU \citep{1981IAUS...93..113H}.

We thank David Latham, Joey Rodriguez, Masahiro Ikoma, Shogo Tachibana, and Hirokazu Kataza for fruitful discussion.  This work is supported
by a Grant-in-Aid from Japan Society for the Promotion of Science (JSPS), Nos. 17K14246, 18H01247, 18H04577, JP20H00170 (H.K.), 17H01103, 18H05441(T.M. \& M.M.), 18K13599 (R.O.), 19K03932 (T.M.), 19H01933 (M.F.), and 14J07182 (M.A.). M.F. is supported by University of Tokyo Excellent Young Researcher Program. M.A. is supported by the Advanced Leading Graduate Course for Photon Science (ALPS). This work was also supported by the JSPS Core-to-Core Program (Planet$^2$) and SATELLITE Research from Astrobiology center (AB022006).
\bibliography{ref}
\begin{figure*}
    \begin{center}
        \includegraphics[width=0.9\linewidth]{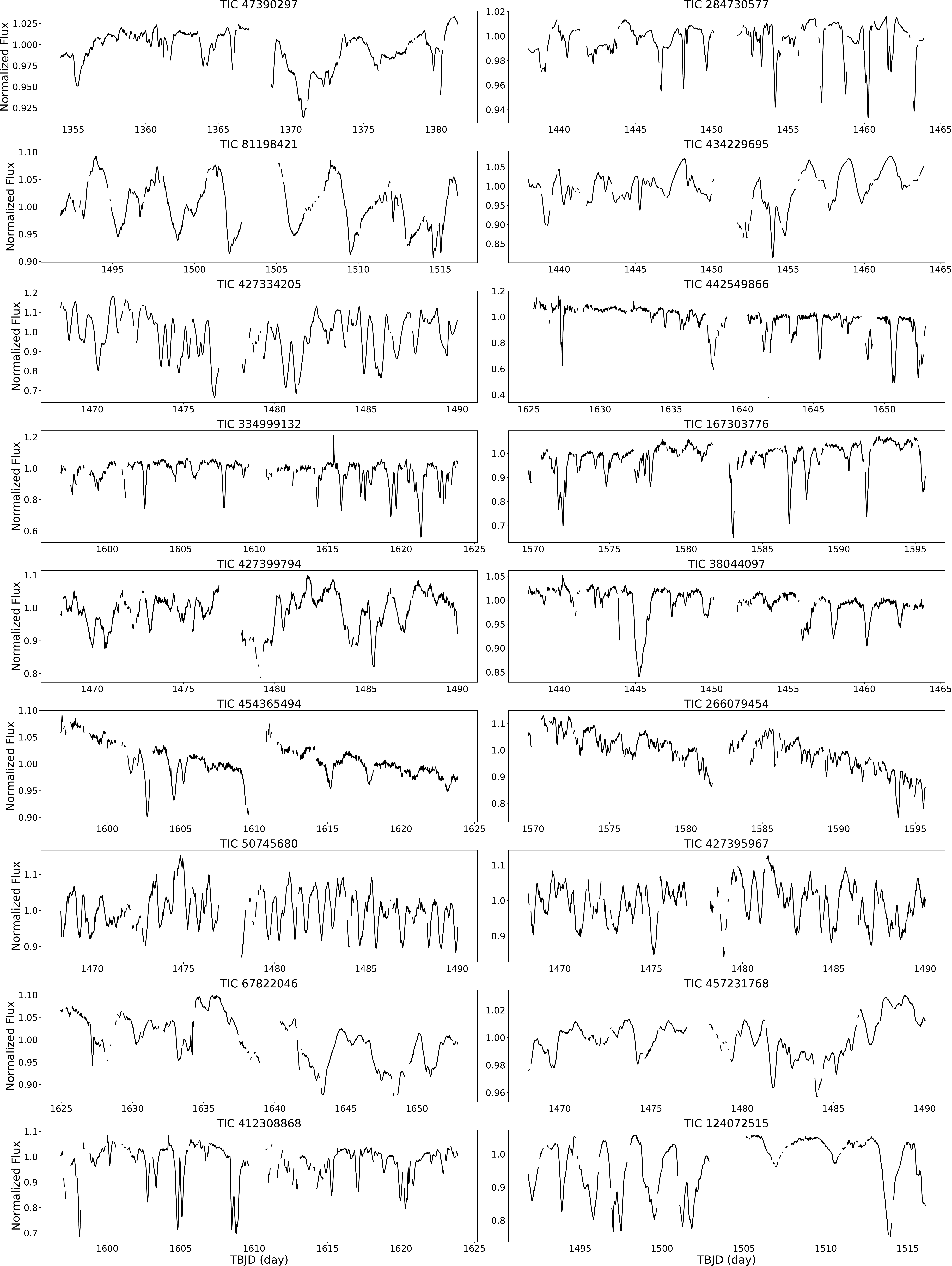}
    \end{center}
\caption{All light curves of dippers we found.}\label{fig:lc1}
\end{figure*}
\begin{figure*}
\addtocounter{figure}{-1}
    \begin{center}
        \includegraphics[width=0.9\linewidth]{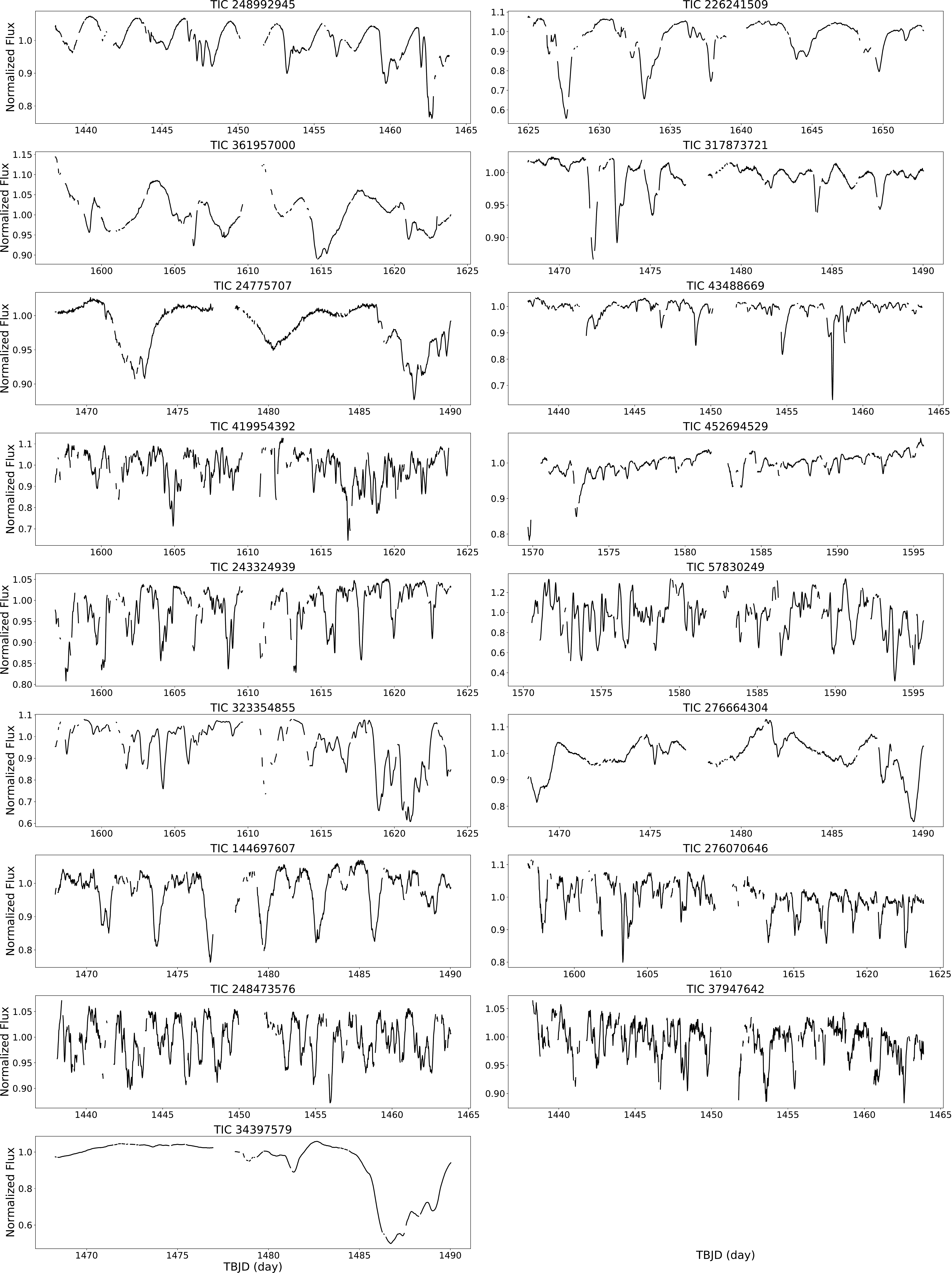}
    \end{center}
\caption{Continued}\label{fig:lc2}
\end{figure*}

\appendix
\section{Reduced Unit Weighted Error}
Our analysis depends on the solution of Gaia data release 2 and TICv8. TICv8 also depends on the resluts of Gaia. The Reduced Unit Weighted Error (RUWE) is often used to distinguish 
single stars from binary or multiple unresolved stars or the targets of Gaia.
RUWE is defined as
\begin{equation}
    RUWE = \frac{\sqrt{\chi \/ \left(N - 5\right)}}{u_0\left(G,C\right)},
\end{equation}
where $\chi$ is {\tt astrometric\_chi2\_al}, $N$ is {\tt astrometric\_n\_good\_obs\_al}, which are listed in the Gaia database, and $u_0\left(G,C\right)$ is a reference value to the normalized Unit Weighted Error, which depends on stellar magnitudes and colors. The $u_0\left(G,C\right)$ is provided by https://github.com/agabrown/gaiadr2-ruwe-tools. If RUWE is close to $1.0$, the source is well-behaved as a single star.
We calculated RUWE for all of the dippers and found that all of RUWE values range between 0.8 and 2 except for TIC 50745680 and 427399794, whose RUWEs are $2.3$ and $10.7$ respectively. This suggests that these stars are not well resolved from nearby stars.

\end{document}